\pdfoutput=1
\documentclass[11pt,twoside,a4paper,cmspaper,final,collab]{cms-tdr}
\def\svnVersion{7aa7628}\def\svnDate{2026/05/23}\def\cmsCernNoTag{CERN-EP-2025-089}\def\cmsCernDate{\today}\def\cmsMessage{Published in the Journal of High Energy Physics as \href{http://dx.doi.org/10.1007/JHEP05(2026)176}{\doi{10.1007/JHEP05(2026)176}.}}
\begin{document}\cmsNoteHeader{HIN-24-007}

\newcommand{\pp}{\ensuremath{\Pp\Pp}\xspace}
\newcommand{\pTjet}{\ensuremath{\pt^{\text{jet}}}\xspace}
\newcommand{\thetal}{\ensuremath$\theta_{\text{l}}$\xspace}
\newcommand{\thetaSD}{\ensuremath$\theta_\mathrm{SD}$\xspace}
\newlength\cmsTabSkip\setlength{\cmsTabSkip}{1ex}

\ifthenelse{\boolean{cms@external}}{\providecommand{\cmsLeft}{upper\xspace}}{\providecommand{\cmsLeft}{left\xspace}}
\ifthenelse{\boolean{cms@external}}{\providecommand{\cmsRight}{lower\xspace}}{\providecommand{\cmsRight}{right\xspace}} 

\providecommand{\cmsTable}[1]{\resizebox{\textwidth}{!}{#1}}

\cmsNoteHeader{HIN-24-007}
\title{Exploring small-angle emissions in charm quark jets in proton-proton collisions at \texorpdfstring{$\sqrt{s} = 5.02\TeV$}{sqrt(s) = 5.02 TeV}}

\author*[cern]{A. Cern Person}

\date{\today}

\abstract{A measurement of the angular structure of inclusive jets and those containing a prompt \PDz meson in proton-proton collisions at the LHC at a center-of-mass energy of 5.02\TeV is presented. The data corresponding to an integrated luminosity of 301\pbinv were collected by the CMS experiment in 2017. Two jet grooming algorithms, late-\kt and soft drop, are used to study the intrajet radiation pattern using iterative Cambridge--Aachen declustering. The splitting-angle distributions of jets with transverse momentum (\pt) of around 100\GeV, obtained with these two algorithms, show that there is a shift of the distribution for jets containing a prompt \PDz meson with respect to inclusive jets. 
The suppression of emissions at small angles observed in the late-\kt grooming approach is consistent with the dead-cone effect, whereas the similar suppression for splittings selected with the soft-drop algorithm appears to be induced by gluon splitting to charm quark-antiquark pairs at large angles. The measured distributions are corrected to the particle level and can be used to constrain model predictions for the substructure of high-\pt charm quark jets.
}

\hypersetup{%
pdfauthor={CMS Collaboration},%
pdftitle={Exploring small-angle emissions in charm quark jets in proton-proton collisions at 5.02 TeV},%
pdfsubject={CMS},%
pdfkeywords={Hadron-Hadron Scattering, Jets and Jet Substructure, Charm Physics}}%

\maketitle 

\section{Introduction}\label{sec:introduction}

The formation of jets, the highly collimated showers of hadrons produced in high-energy collisions, is a rich multiscale process that probes different emergent phenomena of the strong interaction. Jet substructure techniques allow for an understanding of jet shower formation as a function of momentum scale, providing a view into the building blocks of quantum chromodynamics (QCD). 
The radiation pattern of jets is governed by the infrared and collinear divergences of QCD \cite{Dokshitzer:1991wu}. Differences between gluons and light quarks arise from Casimir effects.
For jets initiated by heavy quarks, such as charm or bottom, these divergences are effectively regularized by the quark mass, which leads to formation of jets where most of their momentum is carried by the heavy quarks.
This makes the showering process of heavy-quark jets distinct from those of light-quark or gluon jets.
Thus, measurements of heavy-quark jet substructure are needed in order to improve our understanding of the final-state radiation dependence on a mass scale.

The development of iterative jet declustering techniques has enabled the construction of a proxy for the parton branching process. A recent example is the direct observation of the dead-cone effect in proton-proton ($\Pp\Pp$) collisions at the LHC~\cite{ALICE:2021aqk}. The dead-cone effect in QCD is a suppression of gluon emissions off a heavy quark in a cone of angle $\theta_{\text{d}}=m_{\text{Q}}/E_{\text{Q}}$ around the emitting quark \cite{Dokshitzer:1991fd}, where $m_{\text{Q}}$ and $E_{\text{Q}}$ are the mass and energy of the radiating heavy quark.
In addition, it has been proposed that the dead cone can potentially be used to understand medium-induced radiation in heavy ion collisions, as it provides an opportunity to study a phase-space region dominated by such emission~\cite{Cunqueiro:2022svx, CMS:2024krd, Armesto:2003jh}.
Thus, targeted measurements sensitive to the heavy-quark mass and how it affects the jet shower can provide inputs to improve the description of heavy-flavor jet showers. In this analysis, the primary focus is the substructure of charm quark jets, which are identified by the presence of a prompt \PDz meson contained in a jet, referred to as a \PDz jet in the following. Prompt \PDz mesons are those produced directly from the fragmentation of c quarks, including c quarks from gluon splittings. Both \PDz and its charge-conjugate state \PADz are included in the notation \PDz.

The iterative declustering of a jet allows the kinematic properties and mass effects at the level of an individual emission to be accessed. In an angular-ordered tree, such as the one obtained with the Cambridge--Aachen (CA) pairwise clustering algorithm  \cite{Dokshitzer:1997in, Wobisch:1998wt}, the splittings that are clustered first have the smallest angular separations, where mass effects are the strongest. The dead-cone angle is different for each splitting, since it depends on the energy of the mother branch.

\begin{figure}[ht!]
            \centering
            \includegraphics[width=.46\textwidth]{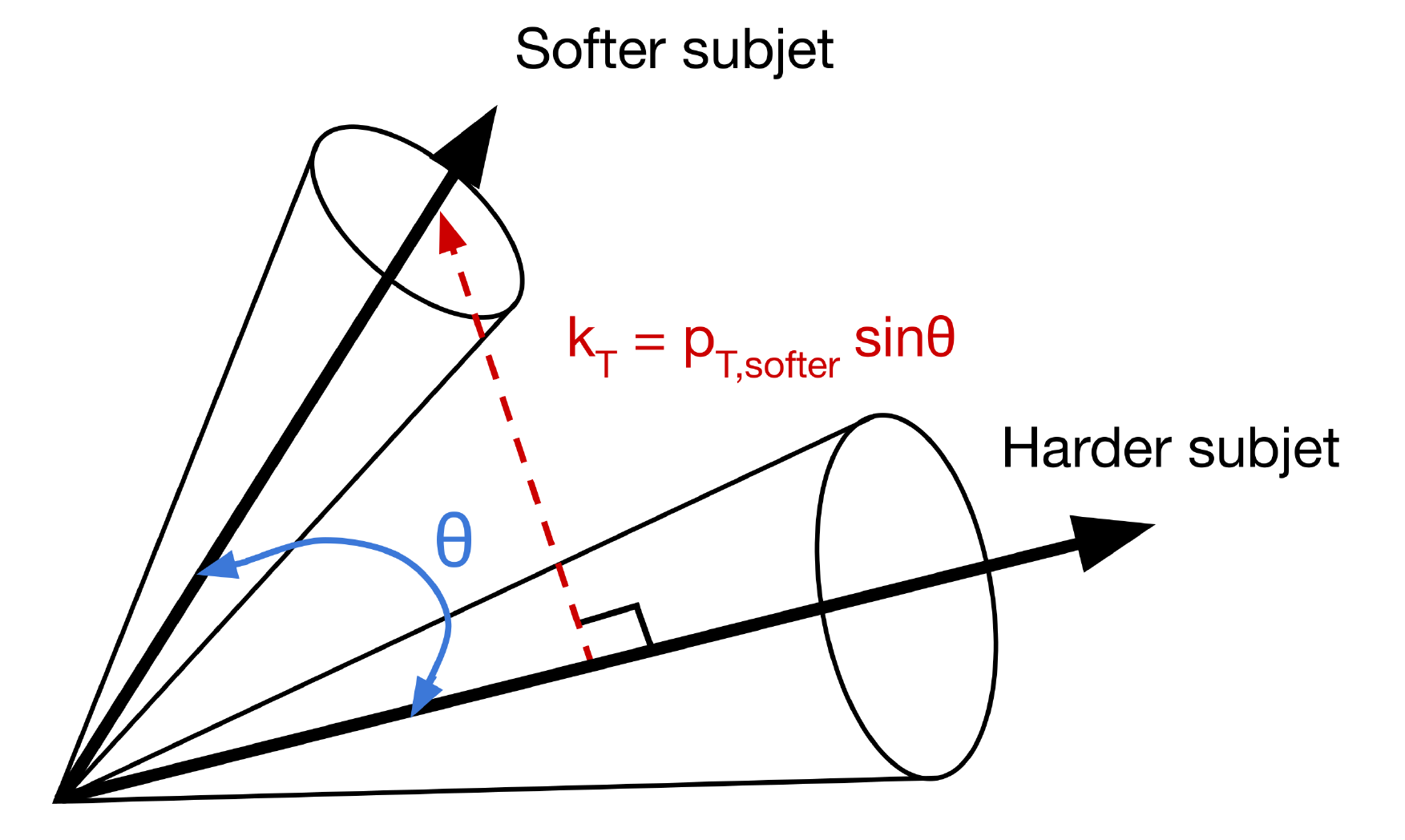}
            \includegraphics[width=.46\textwidth]{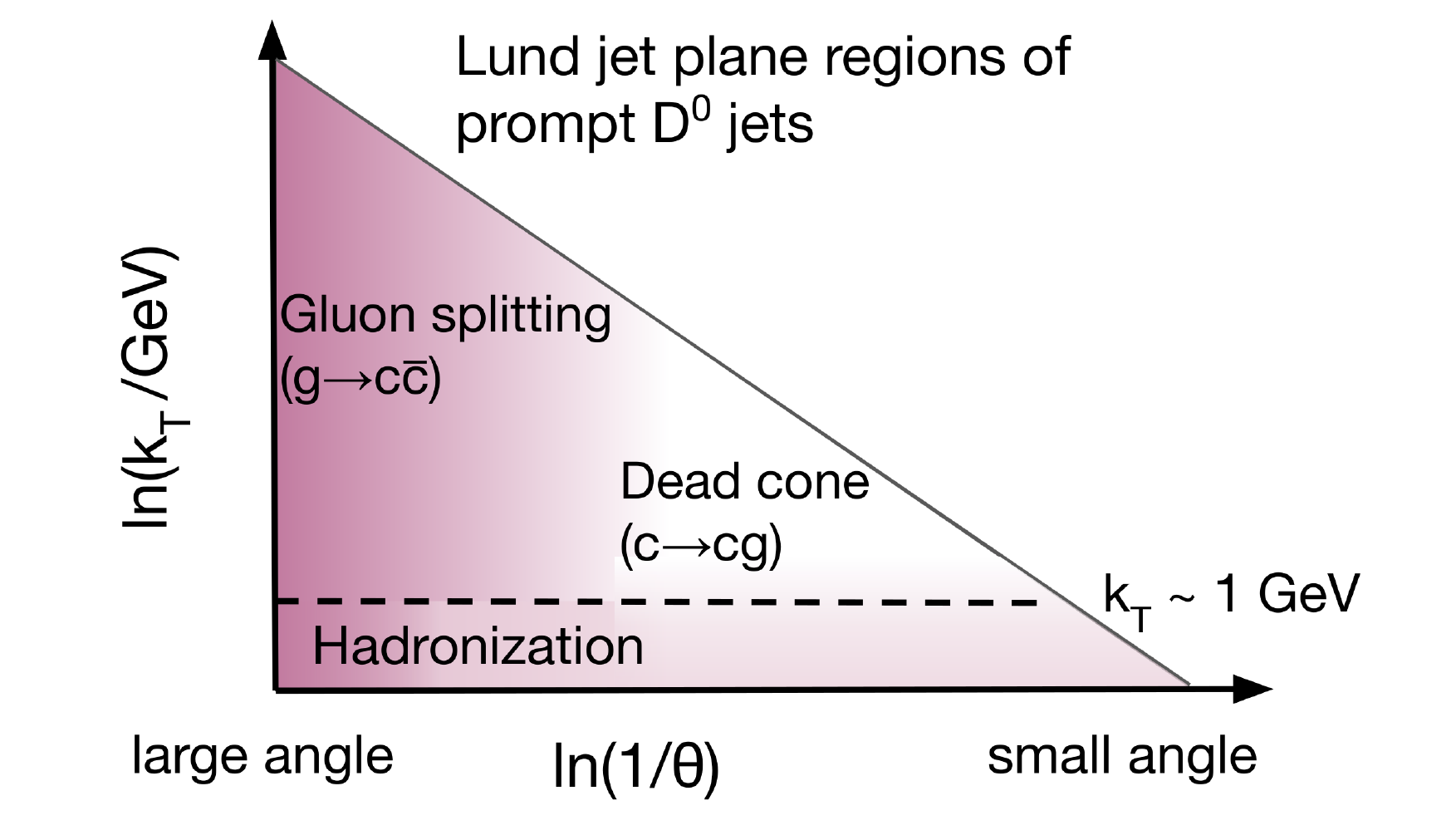}
        \caption{Schematic diagram of two subjets, with their splitting angle $\theta$ and the relative transverse momentum \kt of the softer subjet with respect to the harder subjet (\cmsLeft). Different Lund jet plane regions for charm quark jet showers (\cmsRight), where the vertical axis is the logarithm of the relative momentum \kt of the emission and horizontal axis is the logarithm of the inverse of the angle between the emission and the emitter $\theta$. The shading represents the density of emissions in the primary Lund jet plane for c quark jets.}

        \label{fig:LJP}
    \end{figure}

Using the CA tree, one can map out the kinematics of the intrajet radiation in terms of the Lund jet plane representation \cite{Dreyer:2018nbf}. The Lund jet plane characterizes each individual splitting in terms of the angle $\theta$ between the two resulting jets and the relative transverse momentum of the softer subjet with respect to the harder subjet (\kt). The schematic diagram of a splitting in the left panel of Fig.~\ref{fig:LJP} illustrates these two kinematic quantities. This plane enables the visualization of the jet shower in a modular fashion, as shown in the right panel of Fig.~\ref{fig:LJP}.
For instance, one can suppress the hadronization effects using a selection on \kt, corresponding to a horizontal line in the Lund jet plane. It also allows for isolating angular regions of interest from one another, in particular the regions where different processes producing prompt \PDz jets are located.
Gluon splitting to charm quark-antiquark pairs contributes primarily at large angles $\theta$, hadronization effects contribute primarily to low $\kt < 1\GeV$, and the dead-cone effect is located at small angles.

In Ref.~\cite{Cunqueiro:2018jbh}, two elements were identified as being key for the direct observation of the dead cone:

\begin{itemize}
    \item Suppression of the decays of the heavy-flavor hadrons that can create additional small-angle splittings in the jet clustering tree, such that the dead cone is filled by them. One way to overcome this issue is to select jets with fully reconstructed heavy-flavor hadrons in them.

    \item Suppression of hadronization effects that fill the dead cone. This can be achieved by imposing a \kt cutoff on the splittings that is larger than the hadronization scale.
    \end{itemize}

The ALICE Collaboration used the CA jet tree to obtain direct evidence of the dead-cone effect in jets containing a fully reconstructed \PDz meson \cite{ALICE:2021aqk}.
Here, an alternative strategy is followed to facilitate comparisons with theoretical calculations, reduce nonperturbative corrections and enable exploration of the QCD dead-cone effect in collisions of heavy ions.

In this analysis, the approach consists of selecting one emission (splitting) per jet, such that the mass effects in heavy-flavor jets are exposed and interpretation of the data for both $\Pp\Pp$ and heavy ion collisions is possible, since this measurement will also serve as a reference for future heavy ion studies.
The specific types of emissions are selected by applying a recently proposed grooming algorithm known as ``late-\kt'' \cite{Cunqueiro:2022svx}, designed to select hard and collinear emissions. This algorithm selects the last splitting in the CA jet tree with a value of \kt above 1\GeV, significantly larger than the QCD energy scale $\Lambda_\text{QCD} \approx 250\MeV$ \cite{ParticleDataGroup:2024cfk}. 
This procedure enhances sensitivity to charm quark mass effects while strongly suppressing contributions from hadronization and direct gluon splitting into charm quark-antiquark pairs (g$\to$ \ccbar).
The late-\kt grooming algorithm is applied for the first time in experimental data in this measurement.

A modified version of the soft-drop (SD) grooming algorithm \cite{Larkoski:2014wba,Butterworth:2008iy} with a \kt selection imposed to minimize hadronization effects is also considered.
The emissions selected with the SD algorithm are typically found at larger angles than the ones selected with the late-\kt algorithm. Thus, the SD emissions are potentially more sensitive to effects such as g$\to$ \ccbar or the underlying event activity. Comparing the resulting angular distributions obtained with these two algorithms exposes a modification of the angular scale in \PDz jets relative to inclusive jets in a regime of high transverse momentum (\pt) of the jet. 
In previous measurements of groomed jet substructure of \PDz jets at the LHC~\cite{ALICE:2022phr}, the fragmentation of \PDz jets was constrained at lower jet \pt (\pTjet). 
The dead-cone effect is also studied for the beauty-quark jets by the CMS and LHCb Collaboration~\cite{LHCb:2025mcq,CMS:bjetLund}.

In this analysis, the focus is on moderately high-\pt jets with a \pTjet of at least 100\GeV. This minimum \pTjet allows for a description within the framework of perturbation theory. The ability to penetrate the jet tree and select lower energy splittings within the jet accesses dead cone scales of order $m_{\text{Q}}/E_{\text{splitting}}$, which are significantly higher than $m_{\text{Q}}/$\pTjet, which is the dead cone scale when one considers jets as monolithic objects.
The measurement is performed using $\Pp\Pp$ data at a center-of-mass energy of 5.02\TeV, corresponding to an integrated luminosity of 301\pbinv recorded by the CMS detector in 2017 ~\cite{CMS-PAS-LUM-19-001}. In Section~\ref{sec:cms_detector} the components of the CMS detector are described, as well as the event samples used. Details about the reconstruction of jets and \PDz mesons are presented in Section~\ref{sec:evt_reco}.
Section~\ref{sec:analysis_method} describes the steps in the analysis for both inclusive and prompt \PDz jets, and gives details on the applied corrections.
The strategy for the corrections to stable-particle level is discussed in Section~\ref{sec:corrections}. The systematic uncertainties in the corrected distributions are described in Section~\ref{sec:syst_uncertainties}. The corrected distributions and their comparison with theoretical predictions are reported in Section~\ref{sec:results}. A summary of the measurement is presented in Section~\ref{sec:summary}. Tabulated results are provided in the HEPData record for this analysis ~\cite{hepdata}.

\section{The CMS detector and event samples}\label{sec:cms_detector}

The CMS apparatus~\cite{CMS:2008xjf,CMS:2309.05466} is a multipurpose, nearly hermetic detector, designed to trigger on~\cite{CMS:2020cmk,CMS:2016ngn} and identify electrons, muons, photons, and (charged and neutral) hadrons~\cite{CMS:2020uim,CMS:2018rym,CMS:2014pgm}. A global ``particle-flow'' (PF) algorithm~\cite{CMS:2017yfk} aims to reconstruct all individual particles in an event, combining information provided by the all-silicon inner tracker and by the crystal electromagnetic and brass-scintillator hadron calorimeters, operating inside a 3.8\unit{T} superconducting solenoid, with data from the gas-ionization muon detectors embedded in the flux-return yoke outside the solenoid. The reconstructed particles are used to build \PGt leptons, jets, and missing \pt ~\cite{CMS:2018jrd,CMS:2016lmd,CMS:2019ctu}.

The average number of additional interactions per bunch crossing (pileup) during data taking was two.
This low-pileup activity facilitates the use of neutral hadron PF candidates for the substructure measurement, which eases the comparison of the observable to theoretical calculation.
The events are required to have a primary vertex reconstructed within 25\unit{cm} of the nominal interaction point along the beam direction, and within 2\unit{cm} in the transverse plane.

Events of interest are selected using a two-tiered trigger system. The first level, composed of custom hardware processors, uses information from the calorimeters and muon detectors to select events at a rate of around 100\unit{kHz} for \pp collisions~\cite{CMS:2020cmk}. The second level, known as the high-level trigger, consists of a farm of processors running a version of the full event reconstruction software optimized for fast processing and reduces the event rate to around 1\unit{kHz} before data storage~\cite{CMS:2016ngn}.

Events with high-\pt jets are collected with a combination of triggers requiring at least one jet with energy above 60 and 80\GeV. The trigger has an efficiency of 92\% for offline jets of 80\GeV and of 99\% at 100\GeV.
The trigger effects are simulated in dedicated Monte Carlo (MC) samples. The difference in the trigger efficiency between data and simulation is accounted for in simulated data as a function of \pTjet.

Simulated events used in this analysis are derived at leading order (LO) plus parton shower with the \PYTHIA{} \cite{Sjostrand:2014zea} generator, version 8.230, with the CP5 tune~\cite{CMS:2019csb}, and the \HERWIG{}~\cite{Gieseke:2003rz, Ball:2014uwa, Webber:1983if} generator, version 7.2.2, with the CH3 tune~\cite{Sirunyan:2020pqv}.
These generated samples are passed through a detailed simulation of the CMS detector using \GEANTfour~\cite{Agostinelli:2002hh}.
The \PYTHIA{} sample is used to derive MC corrections, while the \HERWIG{} sample is used to assess systematic uncertainties related to the modeling of the parton shower, hadronization, multiparton interactions, and beam-beam remnant interactions. Additional particle-level predictions are compared with the corrected distributions. These predictions are computed at LO with \PYTHIA{8.303} with the CP5 tune~\cite{Skands:2014pea, Khachatryan:2015pea, CMS:2019csb} and \HERWIG{7} with CH3 tune.

\section{Event reconstruction}\label{sec:evt_reco}

The jets used in this analysis are clustered from the PF candidates provided by the offline event reconstruction using the anti-\kt algorithm \cite{Cacciari:2008gp,Cacciari:2011ma} with a radius parameter $R=0.2$.
The choice of using a small R, compared to the typical 0.4, is convenient since this measurement is intended to be a reference for future planned heavy ion collision measurements where small-$R$ will be used to mitigate the impact of uncorrelated background contributions for jet momentum and substructure \cite{Andrews:2018jcm}.
Moreover, our focus is the small-angle region of the intrajet radiation pattern, which is the most sensitive to quark mass effects.

The four-momentum of the jet is determined using the vector sum of all particle momenta in the jet.
For this analysis, initially selected jets are required to have $80 < \pTjet < 160\GeV$ and pseudorapidity $\abs{\eta} <1.6$. Our results are reported in the region $100 < \pt^\text{jet} < 120\GeV$. A fixed \pTjet window is chosen to limit smearing of the dead-cone region, as the effect depends on the \pTjet. A wider range in $\pt^\text{jet}$ (80--160\GeV) is used for this measurement in order to account for migration when correcting to particle level.
The selection of jets at central $\eta$ is driven by the choice of the \PDz meson candidate selection described in the next paragraph.
Jet energy corrections are derived from independent \pp simulations along with additional corrections, derived using control samples in data, accounting for the imperfect modeling of the detector response \cite{CMS:2011shu}. In addition to the jet energy correction, a correction for the momentum resolution of jets is applied to account for the worse momentum resolution of jets in data compared to simulation. These corrections are derived from dijet balancing studies~\cite{CMS:2016lmd}. To incorporate this effect, a Gaussian smearing is applied to the detector-level \pTjet values in simulation to match the resolution in data.

The \PDz meson candidates are reconstructed by combining pairs of oppositely charged particle tracks with an invariant mass within $\pm 0.2$\GeV of the world-average \PDz meson mass, 1.86\GeV \cite{ParticleDataGroup:2024cfk}. 
To compute the \PDz invariant mass, two hypothesis are considered: 
in the first, one track is assumed to be a pion and the other a kaon; in the second, the particle assignments are swapped.
To suppress the contribution of combinatorial background and to improve the momentum and mass resolution, each track is required to have $\pt>1\GeV$ and $\abs{\eta} <2.4$, and to satisfy a set of track quality selection requirements \cite{CMS:2014pgm}.
In addition, several topological criteria are applied based on the following detector-level variables: the distance between primary and \PDz decay vertices ($d_{0}$) normalized by its uncertainty in the direction transverse to the beam, the angle between the total momentum vector of the daughter tracks and the vector connecting the primary and \PDz decay vertices (pointing angle $\alpha$), the $\chi^2$ probability of \PDz decay vertex fit, and the significance of the distance of the closest approach (DCA) from the total momentum vector to primary vertex.
The selection is optimized using binary trees to maximize the statistical significance of the \PDz meson signal~\cite{TMVA:2007ngy}. The signal input to this optimization is the generator level \PDz, as given by the simulation. 
Each \PDz meson candidate is selected with $\pt > 4\GeV$ and rapidity $\abs{y}<1.2$.
This \PDz \pt requirement is used to suppress the large combinatorial background in \PDz reconstruction in the heavy ion environment, for which the present measurement is used as a baseline \cite{CMS:2019jis}. In addition, it helps to mitigate possible data-to-simulation differences in the reconstruction efficiency of the \PDz meson. Based on MC simulation studies, the \PDz \pt threshold does not introduce a bias in the jet substructure distributions presented in this paper, since the \pTjet used in this measurement is much higher than the \PDz \pt cutoff.

The jets that have at least one \PDz candidate at  $\eta$--$\phi$ distance from the jet axis smaller than the jet radius parameter R, are tagged as \PDz-jets.
Out of the sample of 20 million jets with $100 < \pTjet < 120\GeV$, about 25,000 of them are found to have at least one \PDz meson candidate in data with the characteristics described here.

The observables considered in this analysis are reported at the truth-hadron level. To estimate the impact of hadronization effects, the angular distributions at the parton and at stable-hadron levels using events generated with both \PYTHIA{8} CP5 and \HERWIG{7} CH3 are computed. For a given MC-generated event, the list of partons and stable hadrons is analyzed simultaneously. At the hadron level, prompt \PDz jets are selected with the same kinematic selection requirements on the jets and the \PDz meson in the jet, and with the same grooming algorithm parameters as in the rest of the analysis. At the parton level, jets are required to have a charm quark as a constituent (since there is no \PDz meson formed yet), and the CA reclustering is performed over the list of partons as provided by the respective MC generator. Only parton- and hadron-level jets that are matched geometrically (closest in $\eta$--$\phi$ space, where $\phi$ is azimuthal angle) are used for this comparison. Hadronization effects are quantified through the bin-by-bin difference between the hadron-level jet substructure observable and the substructure of the matched parton-level jet. Hadronization effects are of the order of 10\% across the entire angular range for both late-\kt and SD grooming, demonstrating the resilience of these observables to hadronization effects.

\section{Analysis method}
\label{sec:analysis_method}

This analysis reports the angular distribution of the intrajet emissions in two different regions of the Lund plane for jets containing a prompt \PDz meson and inclusive jets, as selected by two grooming algorithms, the late-\kt and a modified version of the SD algorithm, as described next.

First, the anti-\kt jets~\cite{Cacciari:2008gp,Cacciari:2011ma} are reclustered using the CA algorithm, which is a pairwise clustering algorithm that clusters particles (and subjets thereafter) that are closer in rapidity and azimuth, imposing angular ordering \cite{Dokshitzer:1997in, Wobisch:1998wt}. The PF candidates with $\pt > 1\GeV$ of the original anti-\kt jet are used for the CA reclustering step, which includes neutral and charged PF candidates. 
This \pt threshold is required to ensure high tracking efficiency and reduce data-to-simulation differences.

Then, the CA pairwise clustering history is followed in reverse. Starting with the reclustered jet, the last step of the CA clustering is undone, such that the original jet is ``declustered'' into the two subjets of the previous clustering step. This procedure is performed iteratively, declustering the harder subjet at each step of the iteration. Jet grooming consists of selecting a specific pair of subjets from this iterative CA declustering procedure.

The late-\kt grooming algorithm selects the last pair of subjets that satisfies the $\kt> 1\GeV$ selection, where the angle between the emission and the emitter is defined as the distance in $y$--$\phi$ between the harder and softer subjets in the declustering procedure.
The late-\kt algorithm scans the region of collinear and hard emissions in the Lund jet plane.

The SD grooming algorithm selects the first pair of CA subjets that satisfies the condition $z > z_{\text{cut}} \theta^{\beta}$. Here, $z$ corresponds to the momentum fraction $z=\pt^\text{sub} /(\pt^\text{sub}+\pt^\text{lead})$ where $\pt^\text{lead}$ ($\pt^\text{sub}$) is the momentum of the harder (softer) subjet in the declustering step and $z_{\text{cut}}$ and $\beta$ are SD tunable parameters.
In the modified version of the SD algorithm used in this analysis, the first splitting that satisfies the two requirements is selected:
$z > z_\text{cut} \theta ^{\beta}$ and $\kt>1\GeV$. The parameters are set to $z_\text{cut}=0.1$ and $\beta=0$, for which theoretical calculations can be produced~\cite{Caletti:2023spr}.
The additional \kt requirement is applied to reduce the sensitivity to hadronization effects at small angles, thereby improving the correspondence between the parton- and hadron-level predictions.

For \PDz jets, it is additionally required that the \PDz meson is a constituent of the harder of the two selected subjets.

The angles between the softer and harder subjets selected by the late-\kt or the modified SD grooming algorithm, denoted by \thetal or \thetaSD, respectively, are the main observables reported in this measurement. The present analysis reports such splitting angle distributions for prompt \PDz jets and for inclusive jets. The inclusive jet sample is dominated by light-quark and gluon jets.

\begin{figure}[ht!]
        \centering
        \includegraphics[width=.49\textwidth]{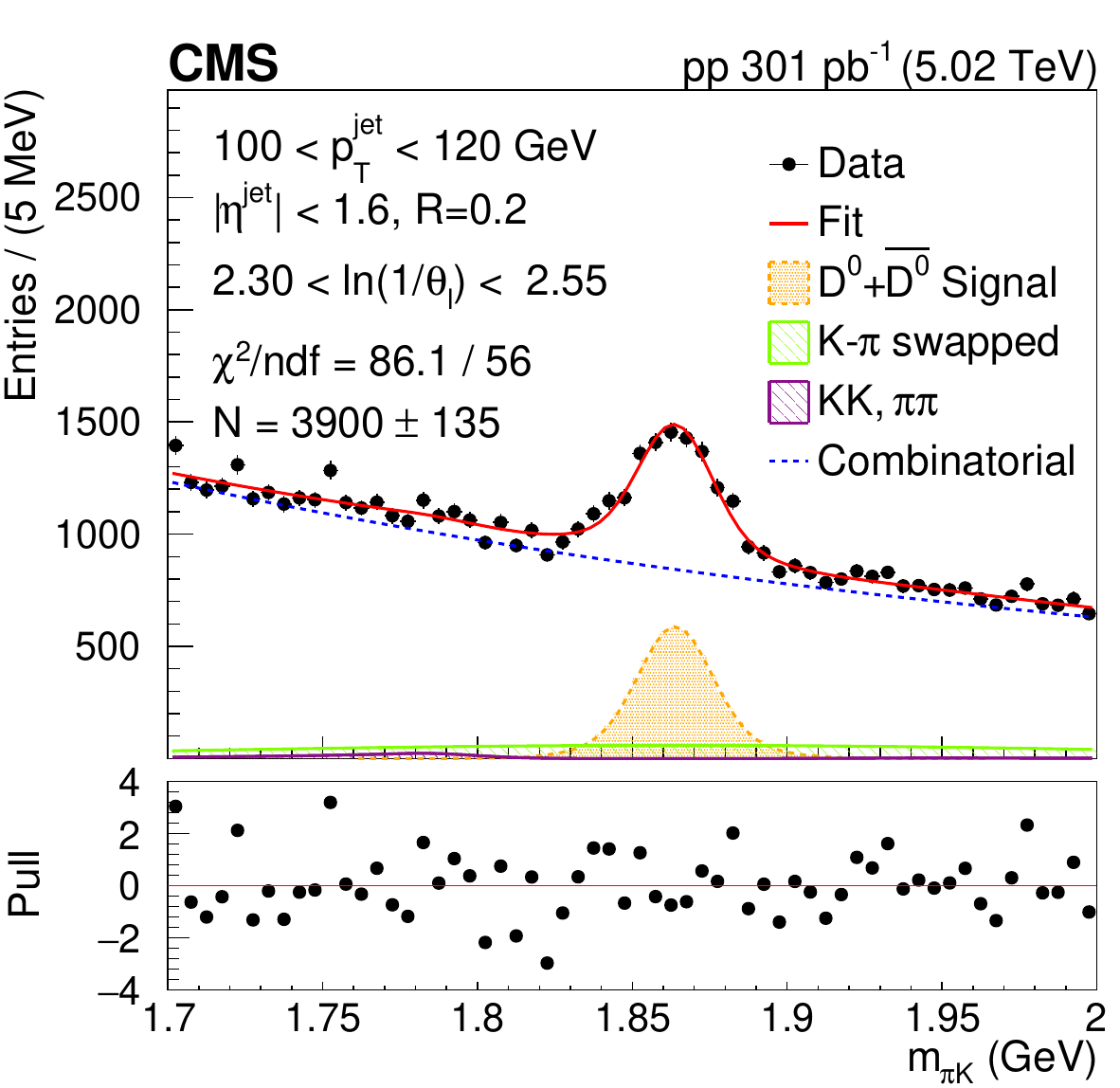}
        \includegraphics[width=.49\textwidth]{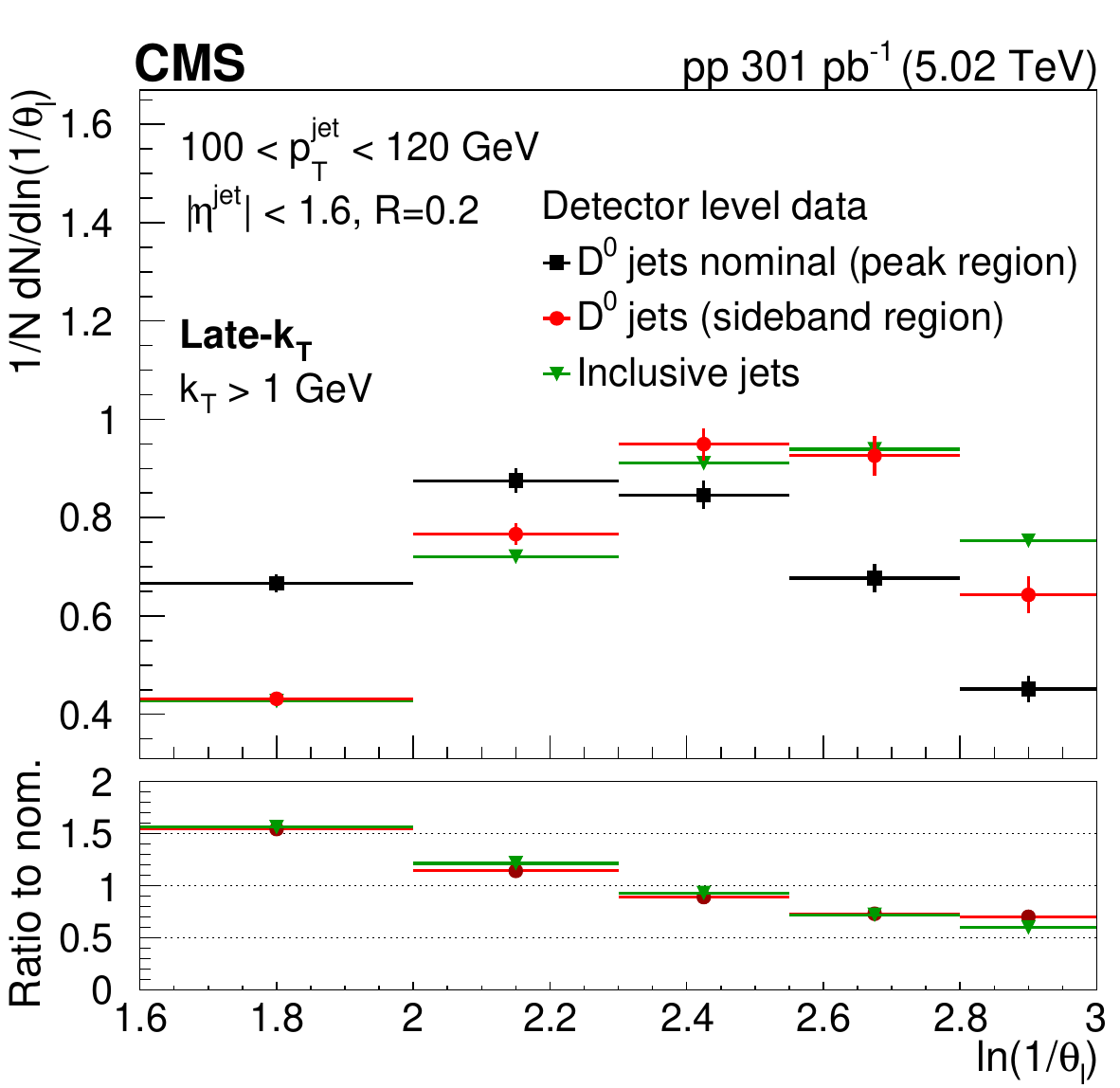}
     \caption{Invariant mass distribution of K$\pi$ pairs for \PDz jet candidates and fits for jets with momentum $100 < \pTjet < 120\GeV$ and late-\kt splitting angle $2.3 <$ \thetal $< 2.55$ (\cmsLeft). Comparison of the ln(1/\thetal) distributions for invariant mass of the track pairs in the resonance region (black rectangles), in the mass sideband region $0.07 < \abs{m_{\pi K} -m^D_{PDG}}<0.12\GeV$ (red circles) and for inclusive jet data (green triangles) (\cmsRight). In the lower panel, a ratio to nominal signal is shown. The error bands represent the statistical uncertainties. The abbreviation ndf stands for the number of degrees of freedom.}
    \label{fig:ExampleFits}
\end{figure}

The analysis workflow for \PDz jets can be summarized by the following steps. The \PDz meson candidates are reconstructed, as described previously. The two charged-particle track daughter candidates are combined to form  the four-momentum of the \PDz meson candidate. This \PDz meson candidate replaces the two daughters, which are removed from the list of particles that are used as input for the jet clustering. The jet substructure observables are computed using the jets that contain at least one \PDz meson candidate as a constituent.
The \PDz meson yield in each \pTjet and \thetal (\thetaSD) bin is extracted with a binned maximum likelihood fit to the invariant mass distributions of the \PDz candidate in the range 1.7--2.0\GeV. The signal shape is modeled as the sum of two Gaussian functions with the same mean but different widths. The combinatorial background originating from random pairs of tracks is modeled with a third-order polynomial function. An additional Gaussian function with a larger width is used to model the invariant mass distribution of \PDz meson candidates that result from the incorrect mass assignment for the pion and kaon tracks. Backgrounds from decays $\PDz \to K^+K^-$ and $\PDz \to \pi^+\pi^-$ are modeled with Crystal Ball functions \cite{Skwarnicki:1986xj}. An example of such fit is shown in the Fig. \ref{fig:ExampleFits} (left).

It is possible that the clustering of a pair of particles with a combined $\pt > 4\GeV$ and with an invariant mass compatible with the \PDz meson could modify the jet substructure distribution so that it mimics the dead-cone effect. In order to test this, a check is performed using the \PDz candidates from low- and high-mass sidebands of the invariant mass distribution \mbox{$(0.07<\abs{m_{\pi K}-m^D_{PDG}}<0.12\GeV)$}. The substructure distribution of sideband-\PDz jets is found to be compatible with the inclusive jet baseline, showing that there are no significant kinematical biases, as seen in the right plot in Fig.~\ref{fig:ExampleFits}.

The yield of \PDz mesons extracted from the invariant mass fits includes both the prompt and nonprompt components.
Nonprompt \PDz mesons are those \PDz mesons that do not originate from the fragmentation of a charm quark but that come from decays of bottom hadrons. 
In this case, the reconstructed jet will contain the \PDz meson, but the fragmentation and dead-cone effects are driven by the mass of the b quark. In addition, it will contain the decay products from the b hadron as constituents. These other decay products may create extra splittings in the CA declustering, leading to contamination in the collinear region of interest.
The prompt signal also includes c resonances, but their contribution is expected to be negligible. For example, as shown in studies by ALICE \cite{ALICE:2021aqk}, $\pi$ mesons coming from D$^*$ decays are very soft, and they contribute to the area below our \kt threshold.

\begin{figure}[ht!]
            \centering
            \includegraphics[width=.49\textwidth]{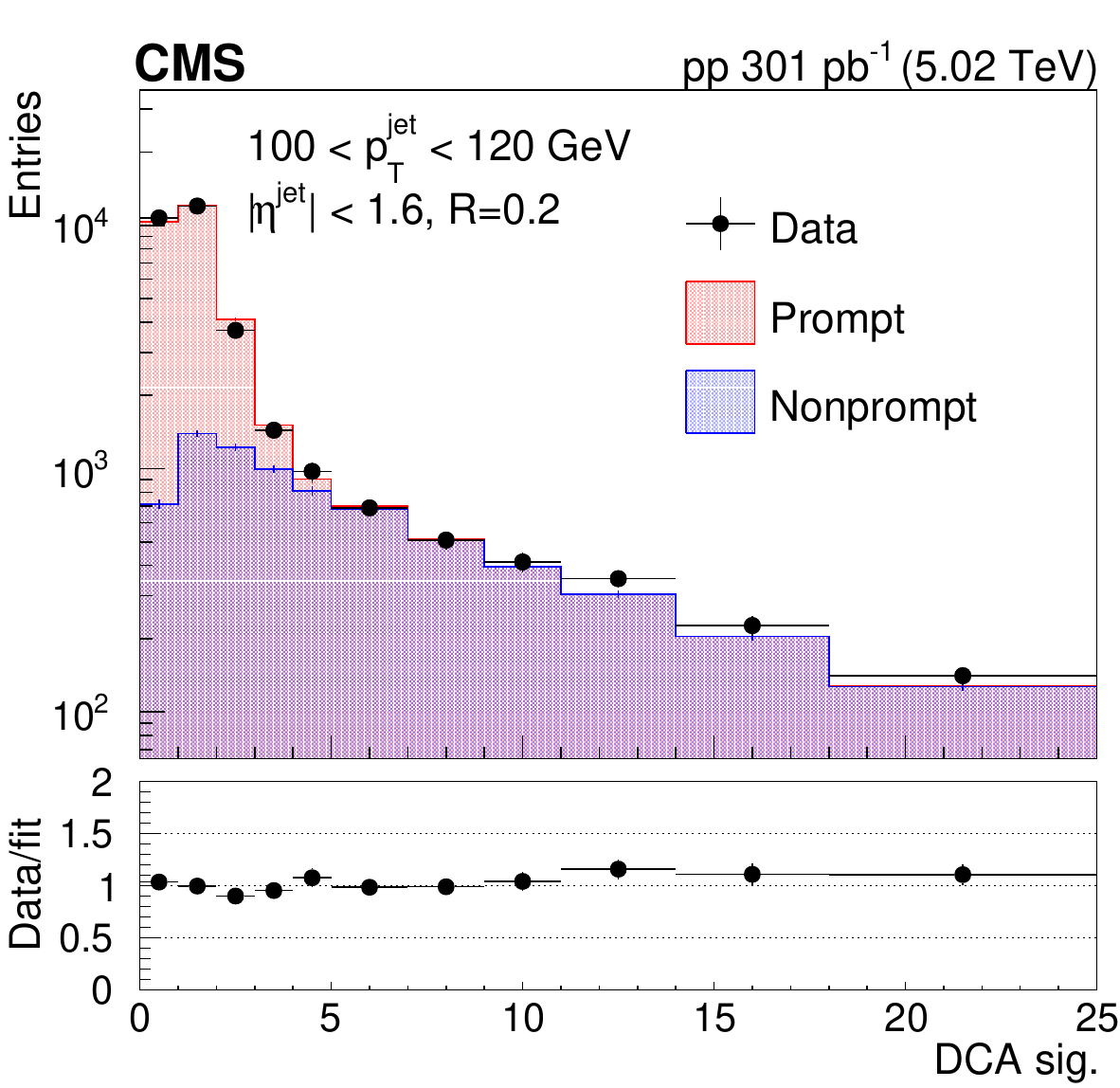}
           \includegraphics[width=.49\textwidth]{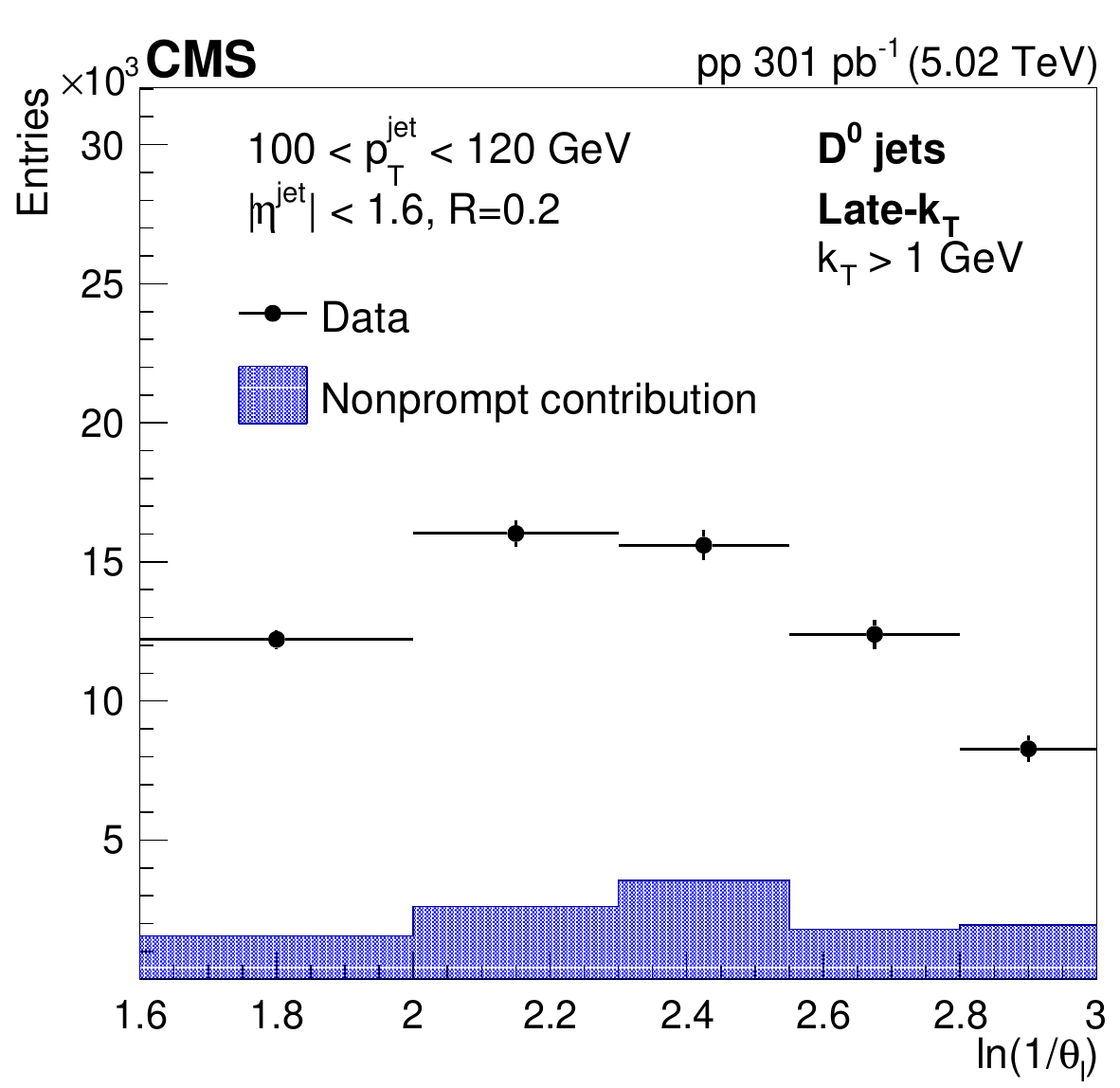}
        \caption{Detector-level DCA significance distribution in data fitted with \PYTHIA{8} CP5 templates for prompt and nonprompt \PDz mesons contained in jets (\cmsLeft). In the lower panel, the ratio between the data and the fit values is presented. Measured \PDz meson yield (black circles) and nonprompt \PDz meson contribution (filled histogram) as functions of the late-\kt splitting angle \thetal (\cmsRight).}
        \label{fig:ExampleNonPrompt}
    \end{figure}

The prompt \PDz contribution is extracted from the DCA significance distribution which shows good separation from the nonprompt component.
Figure~\ref{fig:ExampleNonPrompt} (left) shows the detector-level DCA significance distribution fitted with a linear combination of prompt and nonprompt \PDz DCA templates obtained with \PYTHIA{8} simulated events.
The data distribution, that contains only signal \PDz, is obtained by extracting \PDz yields in bins of DCA significance using the invariant mass fits.

The nonprompt and prompt \PDz meson templates derived from \HERWIG{7} CH3 simulated samples are compatible in shape with the ones derived from \PYTHIA{8}. Figure~\ref{fig:ExampleNonPrompt} (right) shows the \PDz jet substructure distribution and nonprompt \PDz meson contribution obtained with fits in bins of DCA, splitting angle, and \pTjet. The nonprompt \PDz meson contribution is found to be around 15\%.
The nonprompt \PDz subtraction causes a reduction of the uncorrected yield and a change in the shape of the substructure distribution.

To account for bin migration effects due to detector resolution, the measured detector-level distributions are corrected to the particle level using a correction derived from the simulation. The same selections used for data are also applied at the detector-level simulation. To reduce data-to-simulation differences in the jet energy calibration for prompt \PDz jets, the distribution of the ratio of the \PDz and jet momentum in the detector-level simulation is reweighted to match the data distribution. The number of jets that do not satisfy the grooming criteria is tracked throughout the analysis so that all possible migrations are taken into account. Among inclusive jets, approximately 25\% fail grooming criteria, whereas this fraction ranges from 30\% to 40\% for \PDz jets. 
The correction procedure is detailed in the next section.

\section{Corrections to the particle level}
\label{sec:corrections}

The measured detector-level distributions are corrected to the stable-particle level using corrections derived from the simulation. The particle level is defined by the set of particles that have a lifetime longer than $10^{-8}$\unit{s} as given by the MC event generator. For the simulated prompt \PDz jet signal events, the \PDz mesons replace the kaon and pion daughters before the jet reclustering, analogous to what is done at the detector level. The \PDz meson replacement at the particle level uses generator-level information for the mother-daughter association. The same \pt selections on particles are applied at the detector and particle level, including the 4\GeV cutoff on the \PDz meson \pt.
At the particle and detector levels, jets are clustered with a $\pt>1\GeV$ requirement on the particles used for the jet CA reclustering (with the exception of the \PDz meson in the case of the \PDz jet analysis, where an additional selection of \PDz $\pt > 4\GeV$ is required).

The particle-level distribution shares the same binning as the detector-level distribution, and in addition has overflow and underflow bin to account for phase-space migrations in the unfolding procedure.

To derive the corrections, a mapping between the detector- and particle-level emissions is necessary. If the particle- and detector-level jets satisfy the grooming criteria, the splittings selected at the particle and detector levels are required to be, in addition, the closest in proximity among all the primary emissions in the detector- and particle-level CA trees. 

The first step in the correction procedure is to apply a set of bin-by-bin corrections to the detector-level distribution to account for the background (purity corrections). In this measurement, the background consists of detector-level emissions that are not paired with particle-level emissions. Then, to correct for bin-to-bin migrations due to detector effects, a two-dimensional unfolding of \pTjet and the splitting angle (\thetal or \thetaSD) is applied. Unfolding is performed in three bins of \pTjet, where the middle bin is reported, and the surrounding bins are used to allow migrations. Since unfolding with direct matrix inversion yields large bin-to-bin oscillations in the unfolded distributions, the regularized matrix inversion using iterative D'Agostini unfolding with early stopping \cite{DAgostini:1994fjx} is performed. Its implementation in the \textsc{RooUnfold} package \cite{RooUnfold} is used.
After the correction for bin-to-bin migration effects, a set of bin-by-bin corrections to account for particle-level emissions that are not reconstructed at the detector level (efficiency corrections) is applied. Finally, the unfolded result is corrected by the selection, acceptance, and reconstruction efficiency of the prompt \PDz mesons. This correction factor is of the order of 50\% and has a weak dependence on the splitting angle. This emphasizes the fact that decay properties on which our \PDz selection is based are uncorrelated with the jet substructure.

The number of iterations plays the role of the regularization parameter in iterative D'Agostini unfolding. In this analysis, the number of iterations is chosen based on $\chi^2$ goodness-of-fit tests at the detector level. At each iteration of D'Agostini unfolding performed using \PYTHIA{8} CP5, the unfolded distribution is mapped back to the detector level by matrix multiplication using the same migration matrix. 
This process is referred to as ``forward folding,'' and the detector-level distribution obtained after matrix multiplication as the ``forward-folded distribution.'' 
To quantify the compatibility between the forward-folded distribution and the input-measured distribution, the $\chi^2$ of the forward-folded distribution and the measured detector-level distribution are calculated at a given iteration. By construction of the D'Agostini unfolding algorithm, the agreement between the input distribution and the forward-folded distributions improves monotonically at each iteration. To avoid overfitting the unfolded distribution to the statistical fluctuations present in the measured distributions, the unfolding algorithm is stopped at the iteration at which the corresponding $p$-value reaches a plateau. For \PDz jets, the optimal number of iterations is found to be 3, while for inclusive jets the optimal number is 14.

The nominal set of corrections is derived using the sample of \PYTHIA{8} CP5 simulated events. The sample of \HERWIG{7} CH3 simulated events is used to estimate biases in the unfolding corrections and to compute the systematic uncertainties associated with the model dependence of the corrections, as described in Section~\ref{sec:syst_uncertainties}.

The \HERWIG{7} sample is also used to validate the correction procedure performed in this measurement. Detector level distributions from \HERWIG{7} CH3 simulated events are corrected to the stable-particle level using a correction derived from the \PYTHIA{8} CP5 simulated events. The corrected distributions are compared with the stable-particle level distributions from 
\HERWIG{7} and they show good agreement.

\section{Systematic uncertainties}\label{sec:syst_uncertainties}

The experimental uncertainties are propagated by repeating the unfolding procedure with variations of the response matrix, prior distribution, purity, and efficiency corrections. The following systematic uncertainties are considered:

\textit{Jet energy scale and resolution:} The jet energy scale uncertainty is propagated through the unfolding by shifting the \pTjet at the detector level in the simulation according to the $\eta$--\pt dependent jet energy scale uncertainties (3--4\%) ~\cite{CMS:2016lmd}. This uncertainty has an effect through the $80 < \pt^\text{jet} < 160\GeV$ selection requirement.
 The uncertainties in the jet energy resolution measurement are also considered. The \pTjet is further smeared at detector level in simulation to better reproduce the jet energy resolution measured in data. Such a smearing procedure comes with an associated systematic uncertainty (2--4\%), which is propagated through the unfolding procedure.

\textit{PF candidate energy scales:} For a given anti-\kt jet, the four-momenta of the jet constituents are shifted by 1\% for charged particles, 5\% for neutral hadrons, and 3\% for photons~\cite{CMS:2017yfk, CMS:2016lmd, CMS-PAS-JME-10-008}. With these variations, the impact of the individual calibration of PF candidate is estimated. The variations are done independently for each PF candidate type, treating the up and down variations as uncorrelated. This corresponds to six different variations with resulting uncertainty estimates that are added in quadrature.

\textit{Tracking efficiency:} To account for the mismodeling of track reconstruction in the simulation, the track reconstruction efficiency uncertainty is propagated through the unfolding procedure. This is done by randomly removing 3\% of the tracks in the simulation.
This value covers mismodeling of the tracking reconstruction efficiency in the jet core \cite{CMS:2016lmd}.
The tracking efficiency has two effects in this analysis: it affects the reconstruction of \PDz mesons and the substructure observables.
The \PDz reconstruction relies on charged-particle tracks, and the tracking efficiency uncertainty is estimated by removing 3\% of the tracks before the \PDz meson candidate reconstruction. The jet substructure observables are obtained using the PF candidates that have both tracking and calorimeter information. Therefore, rather than losing the PF candidate, it is assumed that the charged hadron is reconstructed as a neutral hadron, and its energy is smeared by 10\% due to the hadronic calorimeter resolution.
The tracking efficiency uncertainty is propagated through the whole chain of corrections in a correlated way, affecting the \PDz meson reconstruction efficiency, the response matrix through the energy resmearing of the charged-particle PF candidates that are ``lost'' in the reconstruction from this variation, as well as the MC-based template fits that are used for the nonprompt \PDz meson background subtraction. The dominant effect is the reduction of the \PDz reconstruction efficiency. This systematic uncertainty is symmetrized bin-by-bin for both \PDz and inclusive jet substructure distributions.

\textit{Physics model uncertainty:} The choice of the physics model impacts the unfolding corrections of the measured observable via the description of the jet showering pattern, which changes the detector response at the subjet level, as well as the prior spectrum used for regularization. The model uncertainty is constructed by using \HERWIG{7} CH3 generated events instead of \PYTHIA{8} CP5 for the prior spectrum, the migration matrix, and the matching purity and efficiency.

\textit{Response matrix statistical uncertainties: } The statistical uncertainties of the simulated sample that is used to derive the migration matrix are propagated through the unfolding procedure, which results in a contribution to the covariance matrix.

\textit{Signal extraction and background modeling:} The \PDz meson yield extraction is performed using a triple Gaussian function, instead of the nominal double Gaussian function used by default. The variation is made to take into account possible deviations in the shape for the nominal configuration. The uncertainty associated with the modeling of the combinatorial background is estimated by using an exponential function and a second-order polynomial instead of the nominal approach of using a third-order polynomial.

\textit{Prompt \PDz meson fraction:} Nominally, the prompt \PDz meson fraction is determined via template fits of prompt and nonprompt \PDz meson production using the DCA significance distribution. To avoid possible detector simulation biases on the DCA significance variable, we use an alternative method for the prompt \PDz meson extraction based only on the DCA distribution. A correction to the detector-level DCA in simulation is applied in order to improve the DCA resolution, following the method described in Ref.~\cite{CMS:2022vfn}. The template fits based on the DCA distribution are repeated after this correction, leading to the uncertainty in the prompt \PDz meson fraction with respect to the nominal DCA significance method.

\textit{Regularization bias:} The optimal number of iterations in the D'Agostini method depends on the initial particle-level spectrum. To gauge the dependence on the initial spectrum, the unfolding of the data using a \PYTHIA{8} CP5 response is repeated for the optimal number of iterations found when unfolding data with a \HERWIG{7} CH3 response. The number of iterations used is 5 and 10, for \PDz jets and inclusive jets, respectively.

\textit{Fit function modeling of the trigger scale factor:} The trigger efficiencies in data and simulation are fitted with an error function parametrization instead of the nominal sigmoid function used as default. The correction chain is repeated with this modification, and the difference relative to the nominal choice is used to quantify this systematic uncertainty.

The resulting relative systematic uncertainties on the normalized jet substructure distributions are summarized in Table ~\ref{tab:Unc_table}. The dominant systematic uncertainties are the jet energy scale, physics model and PF scale uncertainties.

Uncertainties with upward and downward variation, such as the jet energy scale
uncertainty, are symmetrized bin-by-bin. The one-sided variations are kept as one-sided, with the exception of the tracking efficiency uncertainty, which is symmetrized bin-by-bin. The different sources of uncertainty are considered to be independent from each other and their effects are added in quadrature at a given bin. The systematic uncertainties of each source are considered as bin-to-bin fully correlated. The tracking efficiency, PF candidate energy scale, trigger fit modeling, and jet energy resolution uncertainties are propagated as fully correlated bin-by-bin in the \PDz-to-inclusive jet ratio presented in the next section. 
The jet energy scale, physics model uncertainties, and uncertainties specific to the prompt \PDz jet analysis are propagated as uncorrelated in the ratio of \PDz-to-inclusive jet distributions. The statistical correlation between the \PDz jet and the inclusive jet samples is taken into account in the propagation of the statistical uncertainties for the ratio, which results in a statistical correlation coefficient of a few percent.
The total experimental uncertainty in the measurement is obtained by adding in quadrature the total systematic uncertainty and the statistical uncertainty.

\begin{table}[ht!]
\centering
  \topcaption{\label{tab:Unc_table} Summary of fractional uncertainties (\%), with minimum and maximum uncertainty values indicated.}
  \cmsTable{
  \begin{tabular}{lcccc}
    \hline
    \multirow{2}{*}{Uncertainty source} & \multicolumn{2}{c}{Late-\kt} & \multicolumn{2}{c}{SD} \\
    & Prompt \PDz jets & Inclusive jets & Prompt \PDz jets & Inclusive jets \\
    \hline
     Jet energy scale & 1.5--5.8 &  0.3--2.6 & 0.7--4.6 &  0.2--2.4 \\
     Jet energy resolution & 0.3--1.4 & 0.3--0.9 & 0.6--1.5 &  0.2--0.8 \\
     Charged hadron PF energy scale & 0.2--0.6 & 0.7--1.4 & 0.2--0.8 &  0.4--0.8 \\
     Neutral hadron PF energy scale & 1.3--2.8 &  0.3--1.3 &  1.4--5.3 & 0.1--1.5 \\
     Photon PF energy scale & 0.3--1 & 0.5--6 & 0.2--1 &  0.1--0.8 \\
     Tracking efficiency & 0.3--1.2 & 0.1--0.5 & 0.2--2.5 &  0.03--0.3 \\
     Physics model dependence & 0.6--5.4 & 0.4--2.8 & 2.3--7.2 &  0.8--3 \\
     Response matrix statistical & 5.5--9.6 & 0.9--1.5 & 5.2--14.9 &  0.6--1.2 \\
     \PDz Signal extraction & 0.1--0.6 & \NA & 0.02 -0.2 &   \NA \\
     \PDz Background modeling & 0.5--2.8 & \NA & 2.3--7.6 &   \NA \\
     Prompt \PDz fraction & 0.7--3.8 & \NA &  2.4--12.8 & \NA \\
     Regularization bias & 0.4--2.9 & 0.3-0.7 & 0.4--3.5 &  0.1--0.9 \\
     Trigger scale factor & $\leq$ 0.02 & $\leq$ 0.02 & $\leq$ 0.03 &  $\leq$ 0.02 \\
     Statistical & 6.5--11.7 & 1.1--1.7 & 7.1--20 & 0.8--1.4 \\
    \hline
  \end{tabular}
  }

\end{table}

\section{Results}\label{sec:results}

\begin{figure}[ht!]
            \centering
            \includegraphics[width=.49\textwidth]{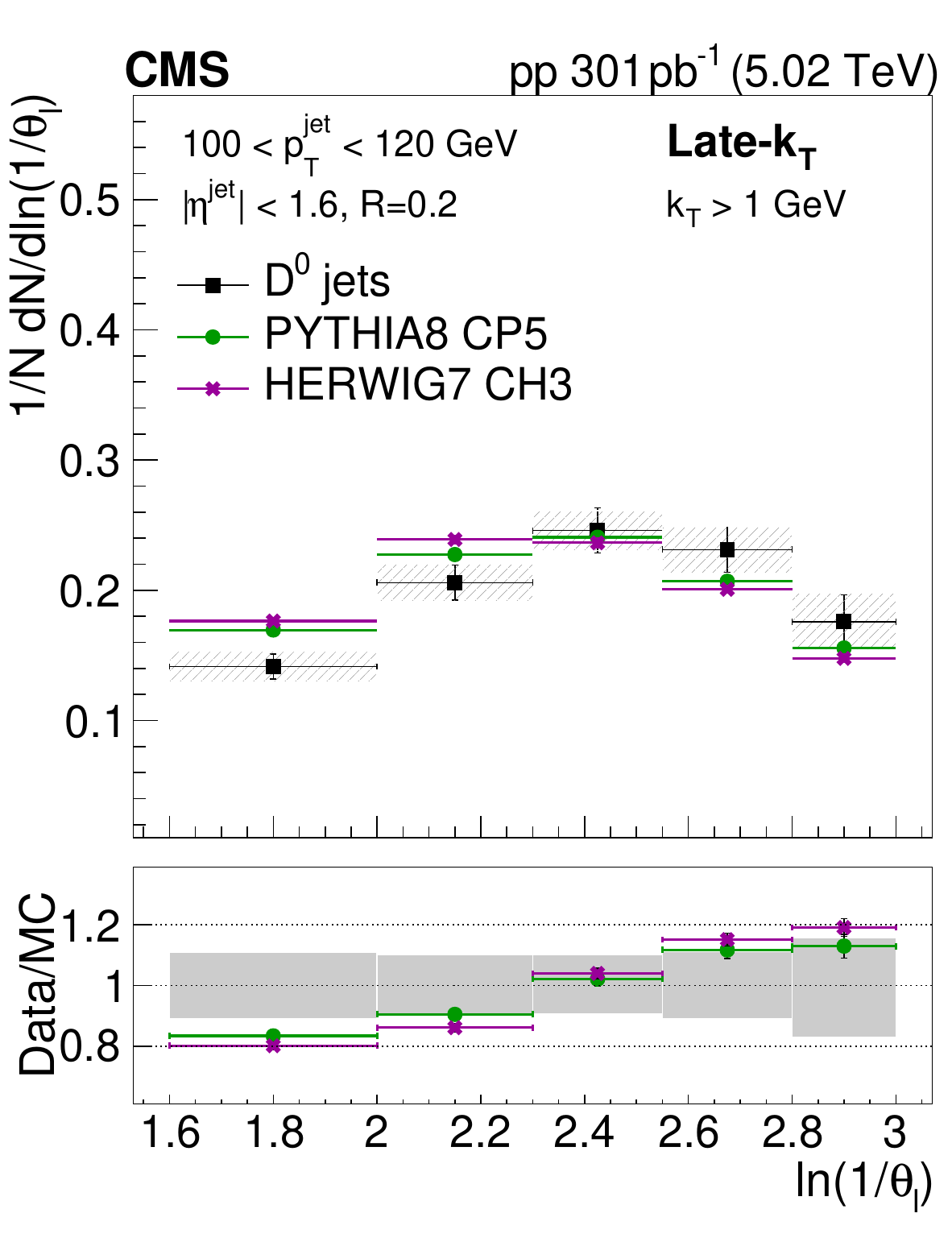}
            \includegraphics[width=.49\textwidth]{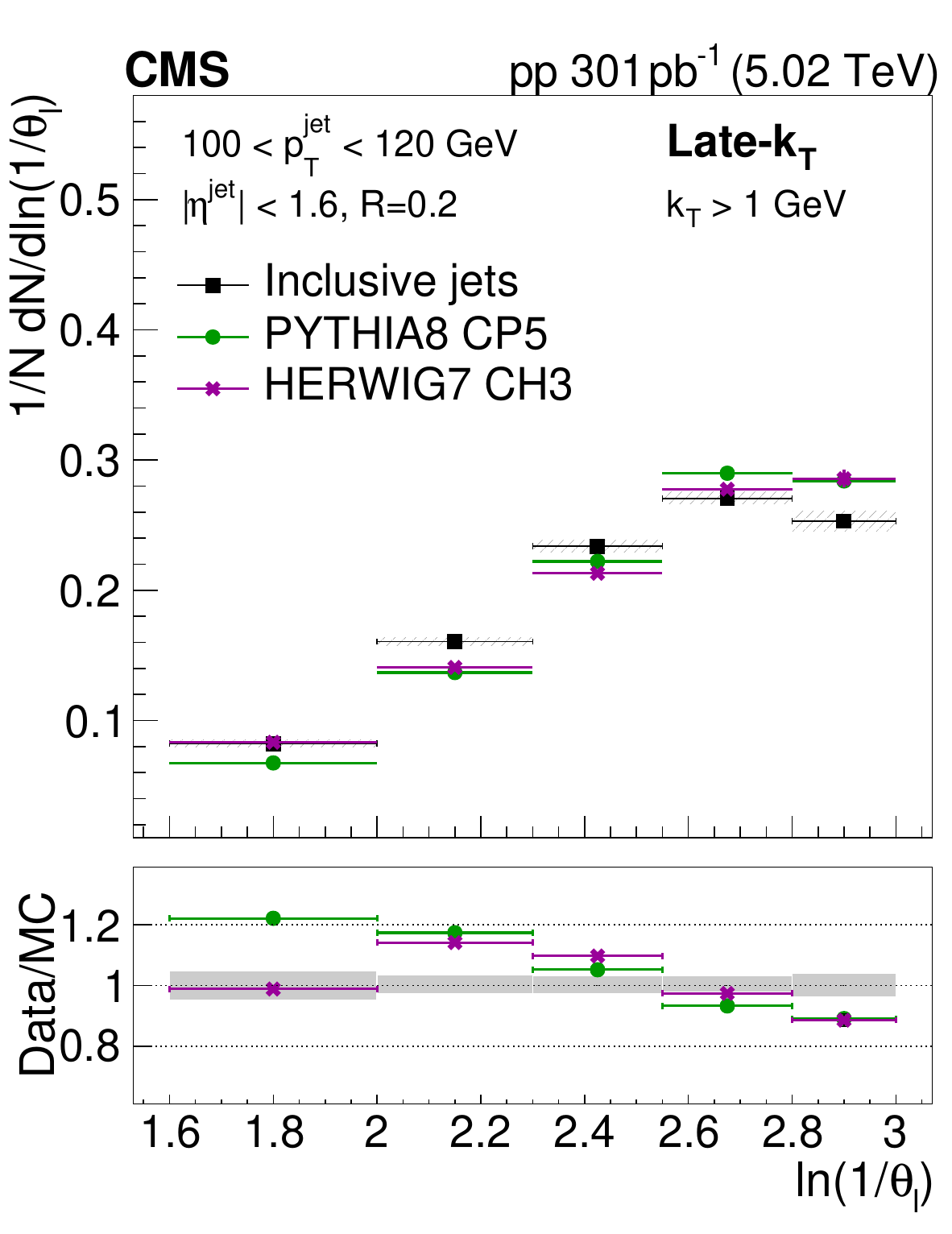}
        \caption{The unfolded late-\kt angular distribution for prompt \PDz jets (\cmsLeft) and inclusive jets (\cmsRight) compared to the predictions from \PYTHIA{8} CP5 and \HERWIG{7} CH3. The error bands in the upper panel represent the total systematical uncertainty, whereas the vertical bars represent the statistical uncertainties. In the lower panel, the error band in the ratio plot represents the total experimental uncertainty in the measurement.}
        \label{fig:Djetsunflatekt}
    \end{figure}

\begin{figure}[ht!]
            \centering
            \includegraphics[width=.49\textwidth]{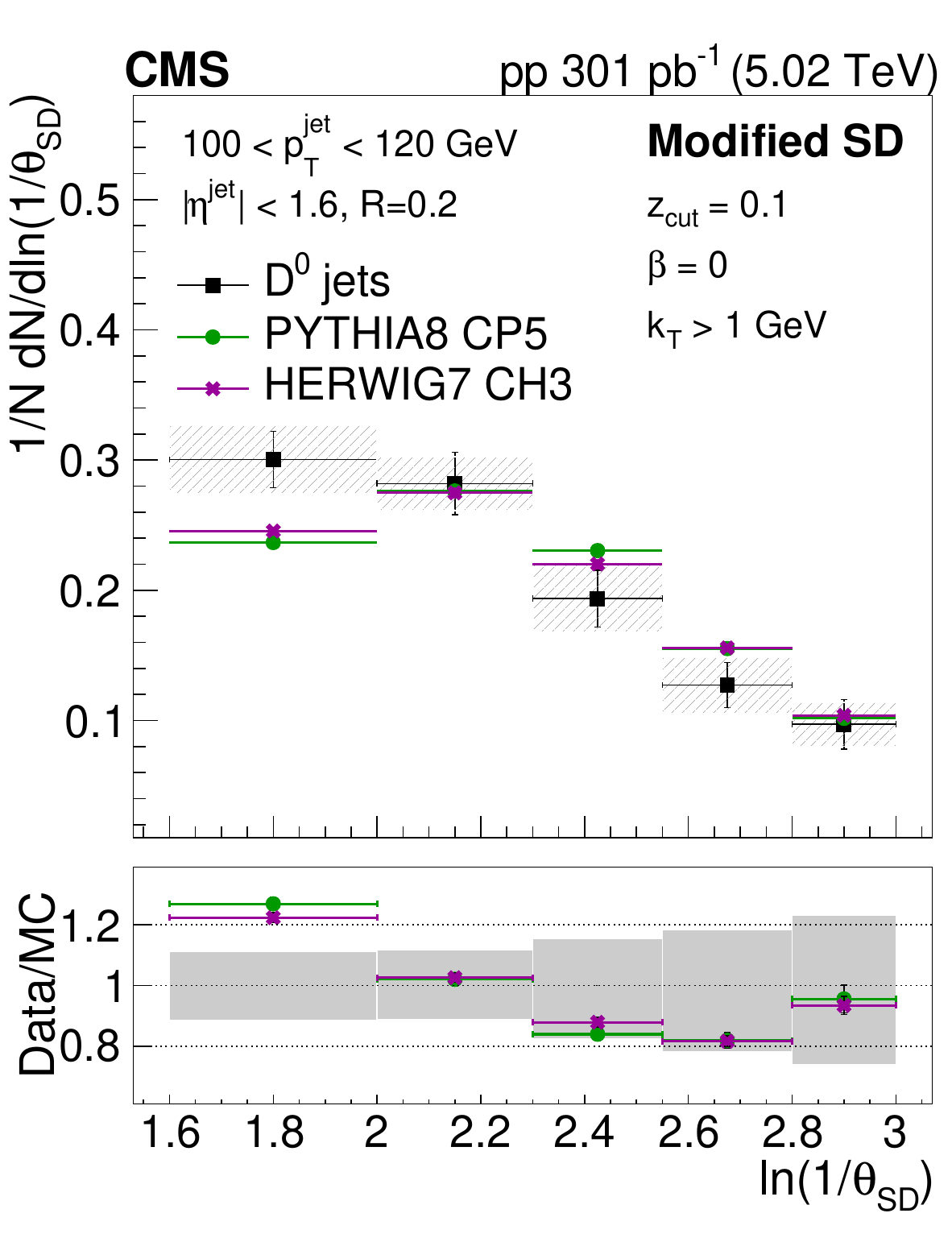}
            \includegraphics[width=.49\textwidth]{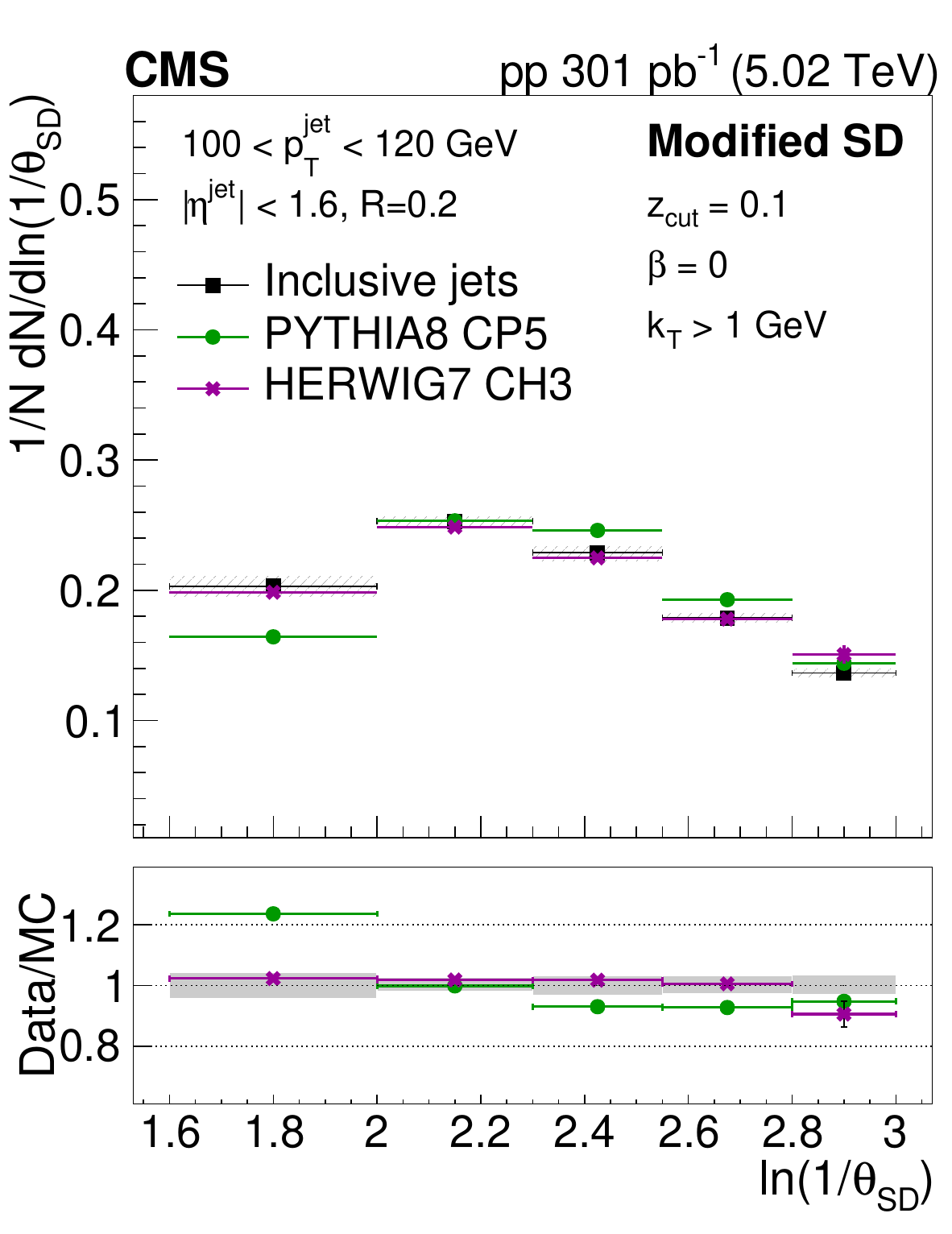}
        \caption{The unfolded SD angular distribution for prompt \PDz jets (\cmsLeft) and inclusive jets (\cmsRight) compared with predictions from \PYTHIA{8} CP5 and \HERWIG{7} CH3. The error bands in the upper panel represent the total systematical uncertainty, whereas the vertical bars represent the statistical uncertainties. In the lower panel, the error band in the ratio plot represents the total experimental uncertainty in the measurement.}
        \label{fig:DjetsunfSD}
    \end{figure}

Figures \ref{fig:Djetsunflatekt} and \ref{fig:DjetsunfSD} present the corrected \thetal and \thetaSD distributions for prompt \PDz jets and inclusive jets. The distributions are normalized to the number of jets that satisfy the respective grooming criteria. The distributions from data are compared to \PYTHIA{8} CP5 and \HERWIG{7} CH3 simulated events at the particle level.
For prompt \PDz jets, the \thetal and \thetaSD distributions from simulated events peak at larger angles compared to the data.
Concerning the distributions for inclusive jets, for the SD grooming algorithm  \HERWIG{7} CH3 describes the data better than \PYTHIA{8} CP5 (Figure~\ref{fig:DjetsunfSD}, right). This is consistent with other jet substructure measurements in inclusive jets~\cite{CMS:2021iwu, CMS:2023lpp} where it is observed that the other tunes with the higher value of \alpS describe data better than CP5, which has a lower cutoff.
The \PYTHIA{8} CP5 and \HERWIG{7} CH3 predictions are consistent with each other for \PDz jets, and the differences between them are larger for inclusive jet production, which is dominated by light-quark and gluon jets.

From Figures \ref{fig:Djetsunflatekt} and \ref{fig:DjetsunfSD}, it can be seen that the splittings selected by the SD grooming algorithm are typically at larger angles compared to those selected by the late-\kt algorithm.

\begin{figure}[ht!]
            \centering
            \includegraphics[width=.49\textwidth]{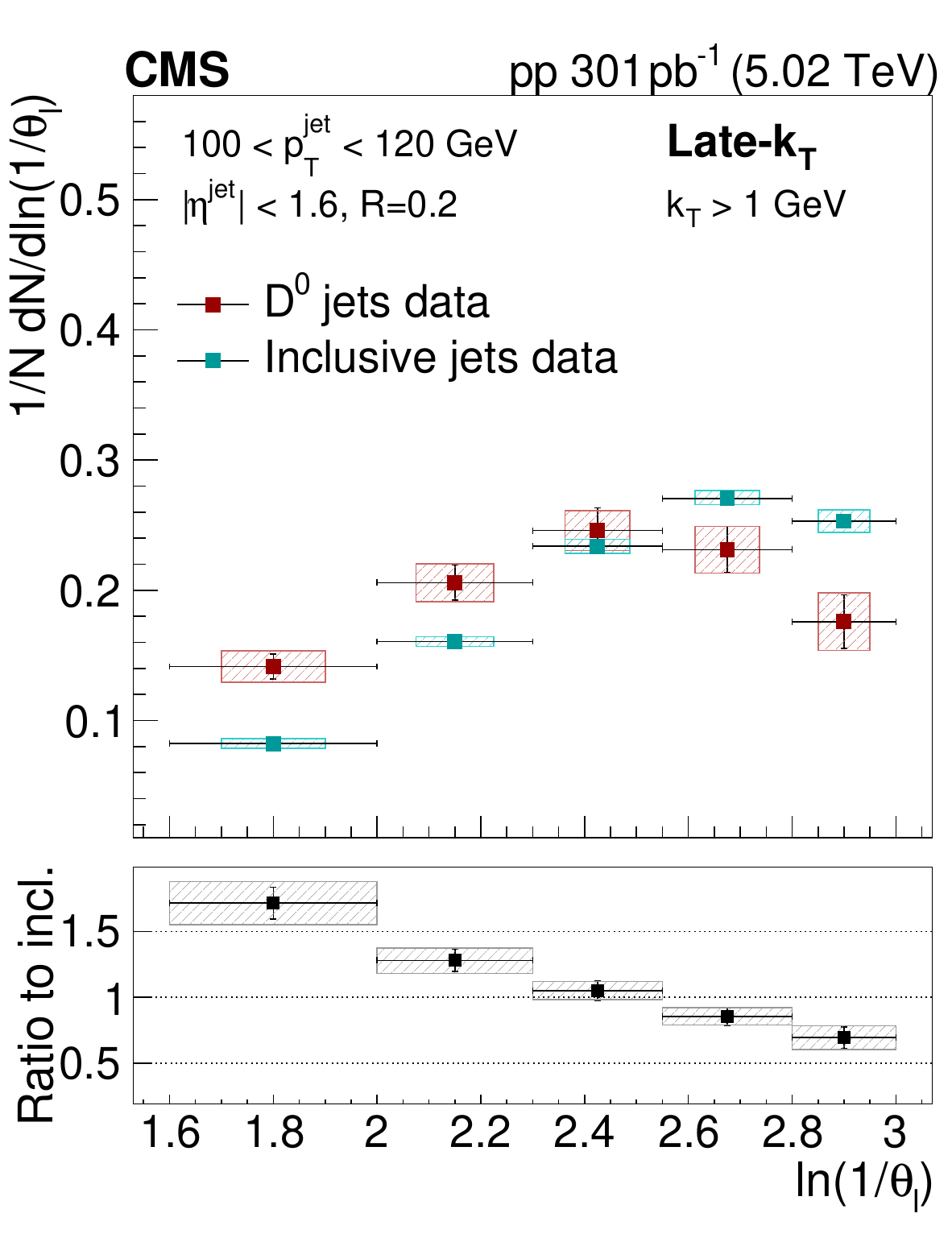}
            \includegraphics[width=.49\textwidth]{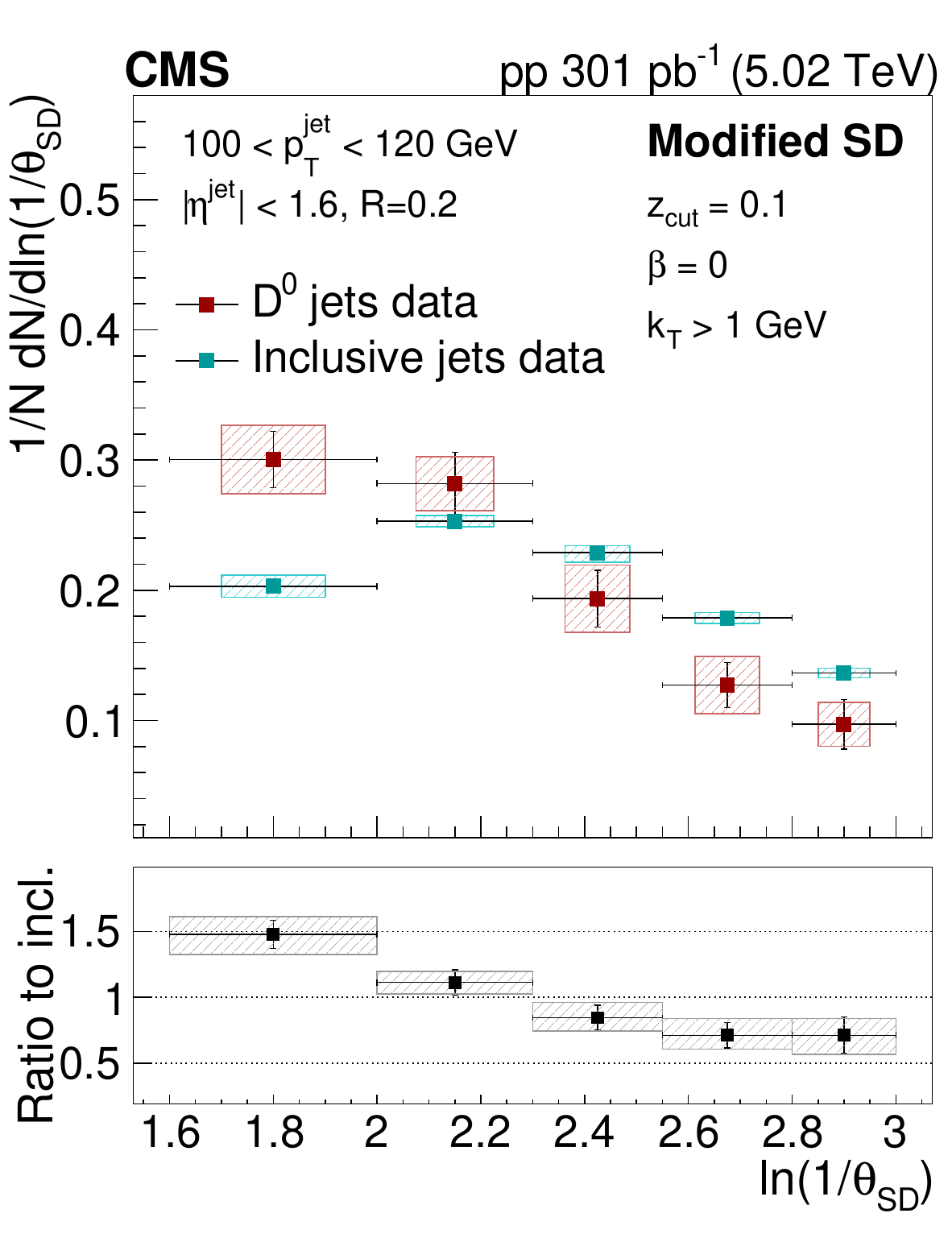}
        \caption{The late-\kt (\cmsLeft) and modified SD (\cmsRight) angular distribution for prompt \PDz jets and inclusive jets. The ratio to the inclusive jets is shown in the lower panels. The error boxes represent the total systematic uncertainty, whereas the vertical bars represent the statistical uncertainties.}
        \label{fig:finalLatekt}
    \end{figure}

Figure~\ref{fig:finalLatekt} presents the \PDz jet and the inclusive jet distributions, as well as the ratio of the \PDz jet distribution to that of inclusive jet. The \PDz jet angular distributions are effectively ``shifted'' towards larger splitting angles relative to the inclusive jet distributions.
Since the QCD dead cone suppresses collinear radiation, the observed shift of emissions towards larger angles is consistent with expectations.

\begin{figure}[ht!]
            \centering
            \includegraphics[width=.49\textwidth]{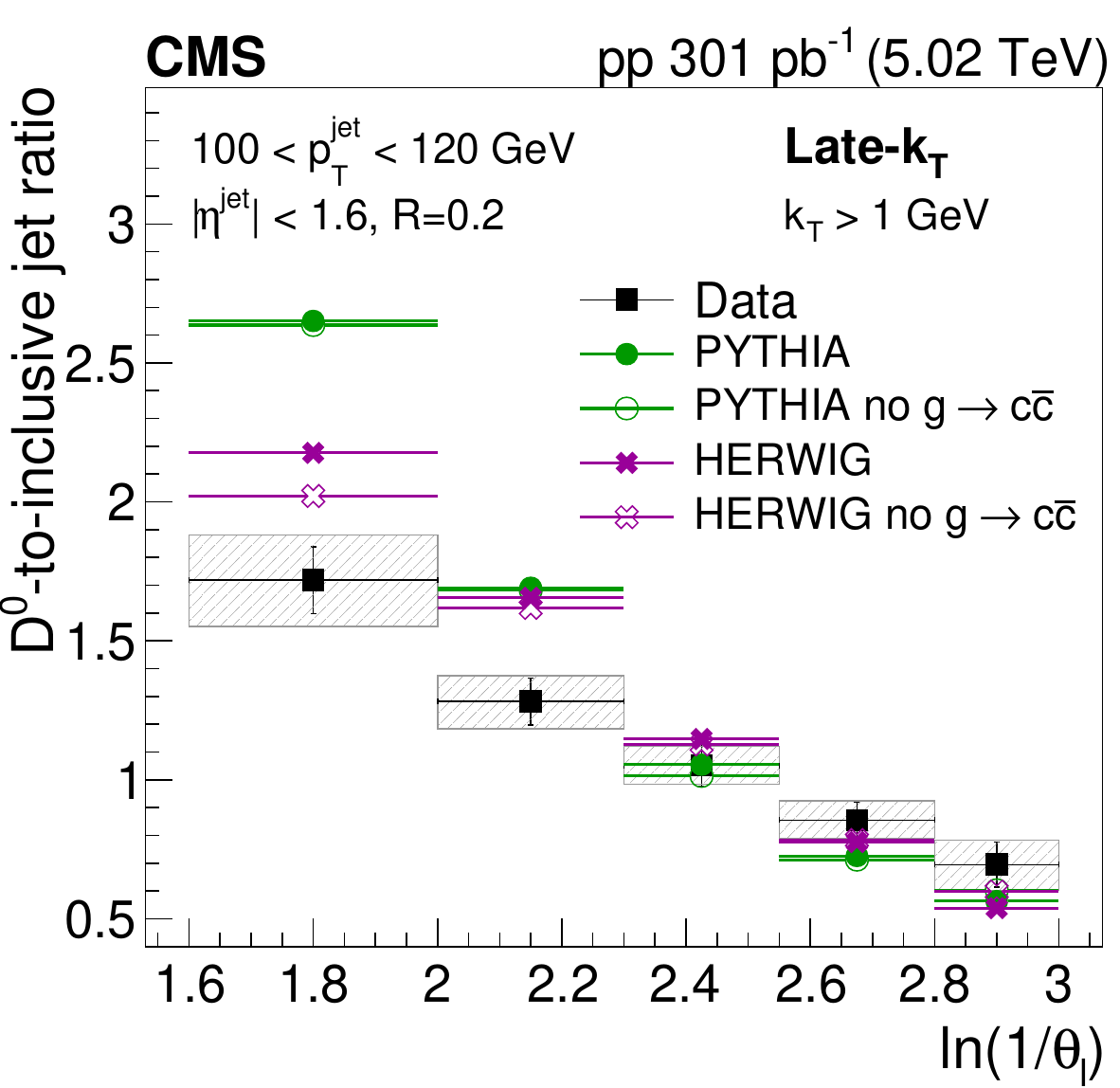}
            \includegraphics[width=.49\textwidth]{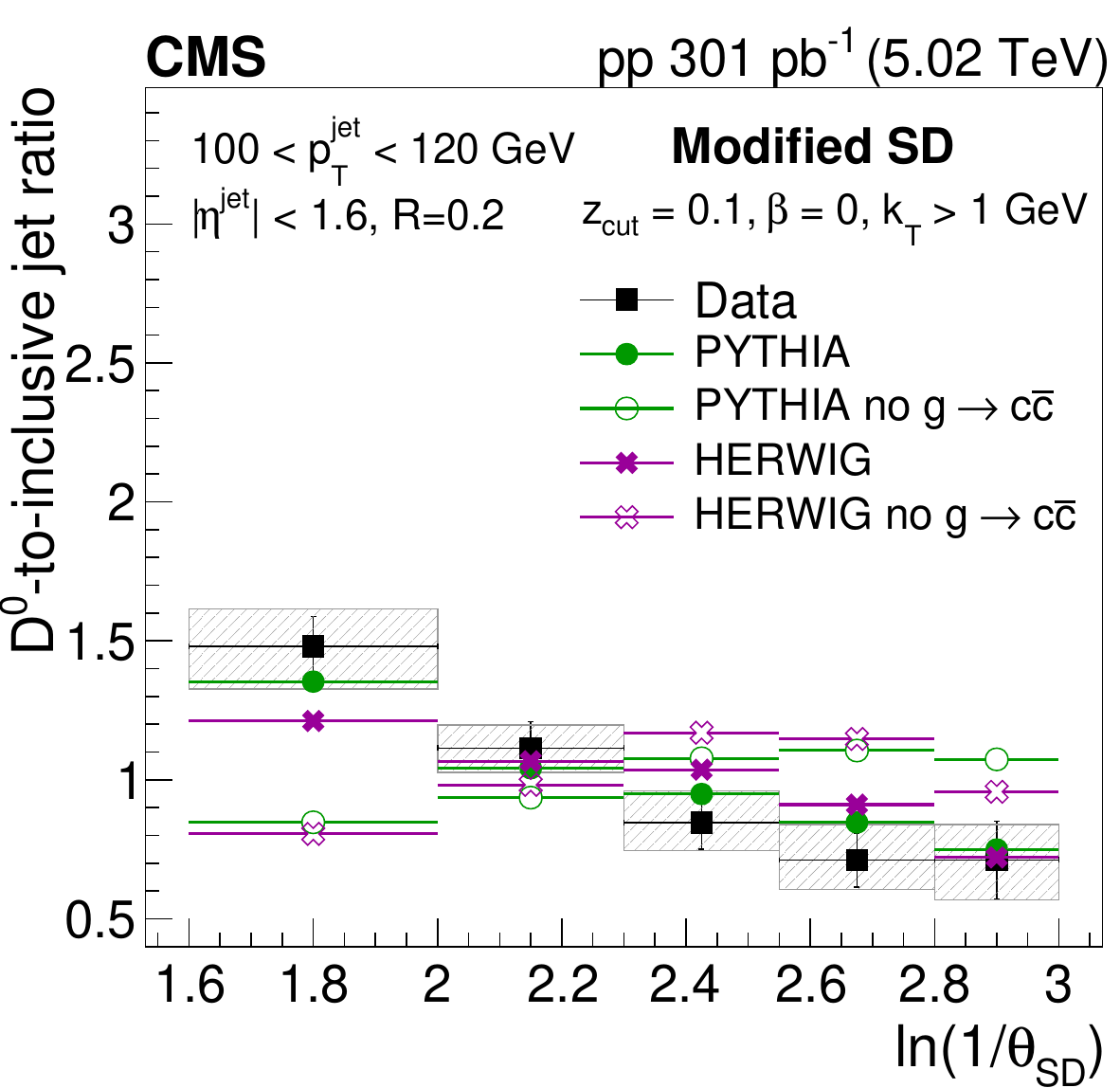}
        \caption{The ratio of the late-\kt (\cmsLeft) and SD (\cmsRight) angle distributions for prompt \PDz jets to inclusive jets. The data are compared to \PYTHIA{8} CP5 and \HERWIG{7} CH3 predictions with and without g $\to$ \ccbar. The error boxes represent the total systematic uncertainty, whereas the vertical bars represent the statistical uncertainties.}
        \label{fig:model_ratio_2a}
    \end{figure}

Prompt \PDz mesons from g$\to$ \ccbar are considered as part of the signal in the analysis. Their impact on the ratio of the distributions of prompt \PDz jets to inclusive jets is studied in \PYTHIA{8} CP5 and \HERWIG{7} CH3 simulated events and shown in Fig.~\ref{fig:model_ratio_2a}.
In both MC generators used in this analysis, events can be generated without gluon splitting to heavy-flavor quark-antiquark pairs, which can be used to understand their contribution to the substructure of prompt \PDz jets. In Fig.~\ref{fig:model_ratio_2a}, gluon splitting is deactivated only for the prompt \PDz jet distributions.
The ratio of \PDz jet to inclusive jet is better described by events simulated with the \HERWIG{7} CH3, which is mostly because \HERWIG{7} CH3 describes the inclusive jet distribution better than \PYTHIA{8} CP5.
For the late-\kt algorithm, the gluon splitting contribution is negligible and has an effect mostly at large angles. For the SD algorithm, the gluon splitting direct contribution is stronger and plays a role in the observed shift between the \PDz jet and inclusive jet distributions. Gluon splitting is found to contribute predominantly at large angles in simulation. The emissions found with SD grooming tend to be at larger angles than the ones found by the late-\kt algorithm, so the contribution of gluon splittings is enhanced for the SD algorithm because of that.

\clearpage

\begin{figure}[ht!]
            \centering
            \includegraphics[width=.49\textwidth]{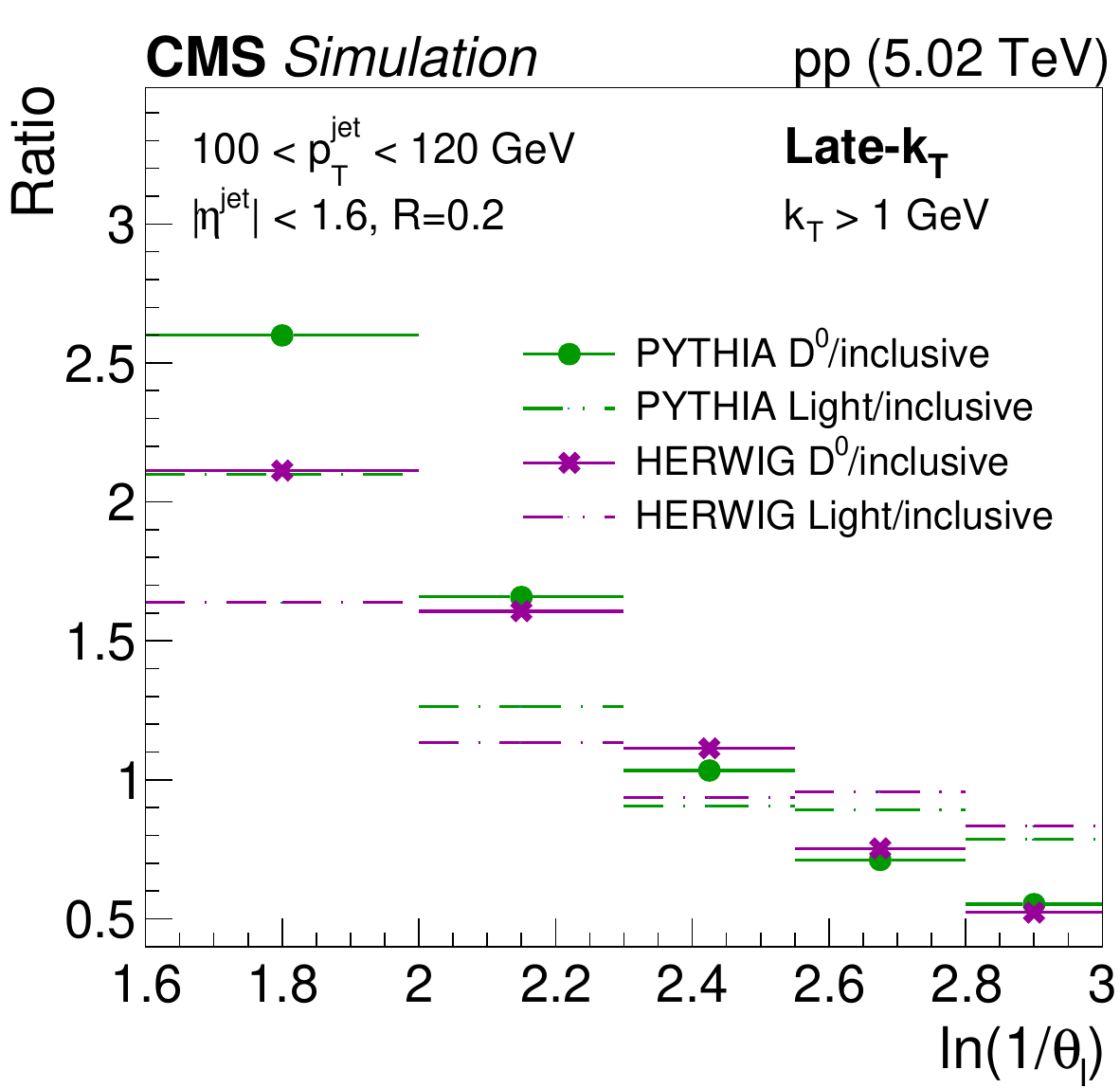}
            \includegraphics[width=.49\textwidth]{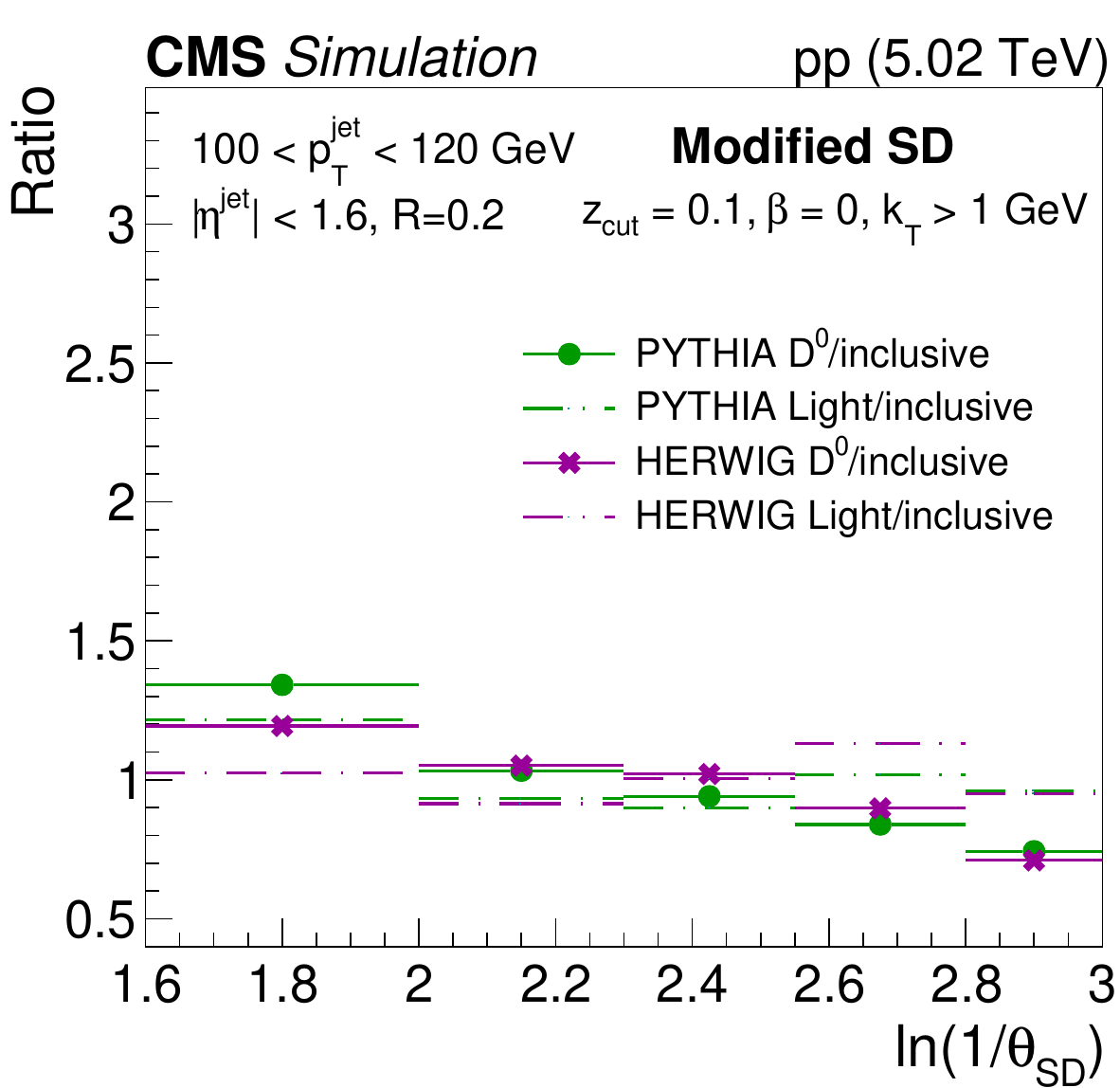}
        \caption{The ratio of the late-\kt (\cmsLeft) and SD (\cmsRight) angle distributions for prompt \PDz jets and light jets to inclusive jets obtained with \PYTHIA{8} CP5 and \HERWIG{7} CH3 simulated events.}
        \label{fig:model_ratio_3}
    \end{figure}

The mass effect in the angular distribution is studied using \PYTHIA{} and \HERWIG{} simulated events. The ratio of the angular distributions of prompt \PDz jets to inclusive jets is compared with the no-dead cone baseline which corresponds to the ratio of the light jets to inclusive jets. From Fig.~\ref{fig:model_ratio_3} it can be seen that these curves have significant differences for both \PYTHIA{} and \HERWIG{} simulated events. Observed differences indicate that for the splittings selected with the late-\kt grooming algorithm, the angular distribution has sensitivity to mass beyond color factor effects.

\section{Summary}\label{sec:summary}

This paper presented measurements of the substructure of jets containing prompt \PDz mesons and of inclusive jets in proton-proton collisions at $\sqrt{s} = 5.02\TeV$, in a data sample corresponding to an integrated luminosity of 301\pbinv, collected in 2017 with the CMS experiment. The analysis focuses on the substructure of jets with transverse momentum $ 100 < \pt^\text{jet} < 120\GeV$ and pseudorapidity $\abs{\eta}< 1.6$, initially clustered with the anti-\kt algorithm and a radius parameter of $R = 0.2$.
Both neutral and charged particles were used for the substructure of these jets. The \PDz mesons were identified via their two-pronged decays into a kaon and pion pair.

In this analysis, the opening angle between the subjet pair found using two different grooming algorithms based on Cambridge--Aachen reclustering was measured. 
The late-\kt algorithm, applied for the first time in experimental data, consists of selecting the last splitting with a $\kt > 1\GeV$ in the Cambridge--Aachen tree and gives access to hard, collinear emissions in an algorithmic way. The angular separation between the two hard subjets found with the soft-drop grooming algorithm was also studied, using the parameters $z_\text{cut} = 0.1$ and $\beta = 0$, with the additional requirement that the emission has a minimum relative transverse momentum of $\kt > 1\GeV$.
Measured angular distributions were compared to \PYTHIA{8} CP5 and \HERWIG{7} CH3 simulated events. \HERWIG{7} CH3 prediction describes angular distribution of inclusive jets better than  \PYTHIA{8} CP5, while in case of jets containing prompt \PDz mesons both predictions fail to reproduce data.
Dedicated \PYTHIA{8} CP5 and \HERWIG{7} CH3 predictions were produced to study impact of the gluon splitting to charm quark-antiquark pairs.
The splitting angle distribution, in the case of soft-drop grooming algorithm, is sensitive to contributions from gluon splitting to charm quark-antiquark pairs at large angles. At the same time, the resulting angular distribution from the late-\kt grooming algorithm is less sensitive to gluon splitting and to soft- and wide-angle radiation.
Although $\pt^\text{jet}$ is much larger than the charm quark mass, it is possible to isolate hard collinear emissions and observe the suppression due to the charm quark mass in the hard and collinear region.

This is the first measurement of charm quark jet substructure for highly energetic jets ($\pt^\text{jet} > 100 \GeV$) that isolates the hard and collinear region of the jet shower, therefore minimizing the hadronization effect and enabling a more direct connection with expected parton shower.
The jet $\pt^\text{jet} > 100\GeV$ selection, used for the first time for charm quark jet substructure, accesses the phase-space region which facilitates the interpretation of the jet substructure in terms of perturbative calculations. 
This measurement will serve as a reference for future studies in heavy ion collisions. In such interactions, the emissions induced by the strongly interacting quark-gluon plasma could be isolated in the region where vacuum emissions are vetoed by the dead cone.

\begin{acknowledgments}
We congratulate our colleagues in the CERN accelerator departments for the excellent performance of the LHC and thank the technical and administrative staffs at CERN and at other CMS institutes for their contributions to the success of the CMS effort. In addition, we gratefully acknowledge the computing centers and personnel of the Worldwide LHC Computing Grid and other centers for delivering so effectively the computing infrastructure essential to our analyses. Finally, we acknowledge the enduring support for the construction and operation of the LHC, the CMS detector, and the supporting computing infrastructure provided by the following funding agencies: SC (Armenia), BMBWF and FWF (Austria); FNRS and FWO (Belgium); CNPq, CAPES, FAPERJ, FAPERGS, and FAPESP (Brazil); MES and BNSF (Bulgaria); CERN; CAS, MoST, and NSFC (China); MINCIENCIAS (Colombia); MSES and CSF (Croatia); RIF (Cyprus); SENESCYT (Ecuador); ERC PRG, RVTT3 and MoER TK202 (Estonia); Academy of Finland, MEC, and HIP (Finland); CEA and CNRS/IN2P3 (France); SRNSF (Georgia); BMBF, DFG, and HGF (Germany); GSRI (Greece); NKFIH (Hungary); DAE and DST (India); IPM (Iran); SFI (Ireland); INFN (Italy); MSIT and NRF (Republic of Korea); MES (Latvia); LMTLT (Lithuania); MOE and UM (Malaysia); BUAP, CINVESTAV, CONACYT, LNS, SEP, and UASLP-FAI (Mexico); MOS (Montenegro); MBIE (New Zealand); PAEC (Pakistan); MES and NSC (Poland); FCT (Portugal);  MESTD (Serbia); MICIU/AEI and PCTI (Spain); MOSTR (Sri Lanka); Swiss Funding Agencies (Switzerland); MST (Taipei); MHESI and NSTDA (Thailand); TUBITAK and TENMAK (T\"{u}rkiye); NASU (Ukraine); STFC (United Kingdom); DOE and NSF (USA).

\hyphenation{Rachada-pisek} Individuals have received support from the Marie-Curie program and the European Research Council and Horizon 2020 Grant, contract Nos.\ 675440, 724704, 752730, 758316, 765710, 824093, 101115353, 101002207, 101001205, and COST Action CA16108 (European Union); the Leventis Foundation; the Alfred P.\ Sloan Foundation; the Alexander von Humboldt Foundation; the Science Committee, project no. 22rl-037 (Armenia); the Fonds pour la Formation \`a la Recherche dans l'Industrie et dans l'Agriculture (FRIA-Belgium); the Beijing Municipal Science \& Technology Commission, No. Z191100007219010, the Fundamental Research Funds for the Central Universities, the Ministry of Science and Technology of China under Grant No. 2023YFA1605804, and the Natural Science Foundation of China under Grant No. 12061141002 (China); the Ministry of Education, Youth and Sports (MEYS) of the Czech Republic; the Shota Rustaveli National Science Foundation, grant FR-22-985 (Georgia); the Deutsche Forschungsgemeinschaft (DFG), among others, under Germany's Excellence Strategy -- EXC 2121 ``Quantum Universe" -- 390833306, and under project number 400140256 - GRK2497; the Hellenic Foundation for Research and Innovation (HFRI), Project Number 2288 (Greece); the Hungarian Academy of Sciences, the New National Excellence Program - \'UNKP, the NKFIH research grants K 131991, K 133046, K 138136, K 143460, K 143477, K 146913, K 146914, K 147048, 2020-2.2.1-ED-2021-00181, TKP2021-NKTA-64, and 2021-4.1.2-NEMZ\_KI-2024-00036 (Hungary); the Council of Science and Industrial Research, India; ICSC -- National Research Center for High Performance Computing, Big Data and Quantum Computing, FAIR -- Future Artificial Intelligence Research, and CUP I53D23001070006 (Mission 4 Component 1), funded by the NextGenerationEU program (Italy); the Latvian Council of Science; the Ministry of Education and Science, project no. 2022/WK/14, and the National Science Center, contracts Opus 2021/41/B/ST2/01369 and 2021/43/B/ST2/01552 (Poland); the Funda\c{c}\~ao para a Ci\^encia e a Tecnologia, grant CEECIND/01334/2018 (Portugal); the National Priorities Research Program by Qatar National Research Fund; MICIU/AEI/10.13039/501100011033, ERDF/EU, "European Union NextGenerationEU/PRTR", and Programa Severo Ochoa del Principado de Asturias (Spain); the Chulalongkorn Academic into Its 2nd Century Project Advancement Project, and the National Science, Research and Innovation Fund via the Program Management Unit for Human Resources \& Institutional Development, Research and Innovation, grant B39G680009 (Thailand); the Kavli Foundation; the Nvidia Corporation; the SuperMicro Corporation; the Welch Foundation, contract C-1845; and the Weston Havens Foundation (USA).
\end{acknowledgments}

\bibliography{auto_generated} 

\cleardoublepage \appendix\section{The CMS Collaboration \label{app:collab}}\begin{sloppypar}\hyphenpenalty=5000\widowpenalty=500\clubpenalty=5000
\cmsinstitute{Yerevan Physics Institute, Yerevan, Armenia}
{\tolerance=6000
V.~Chekhovsky, A.~Hayrapetyan, V.~Makarenko\cmsorcid{0000-0002-8406-8605}, A.~Tumasyan\cmsAuthorMark{1}\cmsorcid{0009-0000-0684-6742}
\par}
\cmsinstitute{Institut f\"{u}r Hochenergiephysik, Vienna, Austria}
{\tolerance=6000
W.~Adam\cmsorcid{0000-0001-9099-4341}, J.W.~Andrejkovic, L.~Benato\cmsorcid{0000-0001-5135-7489}, T.~Bergauer\cmsorcid{0000-0002-5786-0293}, K.~Damanakis\cmsorcid{0000-0001-5389-2872}, M.~Dragicevic\cmsorcid{0000-0003-1967-6783}, C.~Giordano, P.S.~Hussain\cmsorcid{0000-0002-4825-5278}, M.~Jeitler\cmsAuthorMark{2}\cmsorcid{0000-0002-5141-9560}, N.~Krammer\cmsorcid{0000-0002-0548-0985}, A.~Li\cmsorcid{0000-0002-4547-116X}, D.~Liko\cmsorcid{0000-0002-3380-473X}, I.~Mikulec\cmsorcid{0000-0003-0385-2746}, J.~Schieck\cmsAuthorMark{2}\cmsorcid{0000-0002-1058-8093}, R.~Sch\"{o}fbeck\cmsAuthorMark{2}\cmsorcid{0000-0002-2332-8784}, D.~Schwarz\cmsorcid{0000-0002-3821-7331}, M.~Sonawane\cmsorcid{0000-0003-0510-7010}, W.~Waltenberger\cmsorcid{0000-0002-6215-7228}, C.-E.~Wulz\cmsAuthorMark{2}\cmsorcid{0000-0001-9226-5812}
% ...
\par}
\vskip\cmsinstskip
\dag:~Deceased\\
$^{1}$Also at Yerevan State University, Yerevan, Armenia\\
$^{2}$Also at TU Wien, Vienna, Austria\\
% ...
\end{sloppypar}
\end{document}